\documentclass[a4paper,UKenglish,cleveref,autoref,thm-restate]{lipics-v2021}

\hideLIPIcs  

\title{Beyond FLOPs: Energy-Aware Knowledge Distillation for Sustainable LLMs on Code-Related Task} 

\titlerunning{Energy-Aware Knowledge Distillation for Sustainable LLMs on Code-Related Task} 

\author{Enrique {Barba Roque}}{Delft University of Technology, Delft, The Netherlands \and \url{https://ebarba.com}}{enrique@ebarba.com}{https://orcid.org/0000-0002-7018-498X}{}

\author{Luís {Cruz}}{Delft University of Technology, Delft, The Netherlands}{l.cruz@tudelft.nl}{https://orcid.org/0000-0002-1615-355X}{}

\author{Annibale {Panichella}}{Delft University of Technology, Delft, The Netherlands}{a.panichella@tudelft.nl}{https://orcid.org/0000-0002-7395-3588}{}

\authorrunning{E. Barba Roque, L. Cruz and A. Panichella} 

\Copyright{Enrique Barba Roque and Luís Cruz and Annibale Panichella} 



\category{Technical Track Paper} 

\relatedversion{} 

\nolinenumbers 

\EventEditors{Robert Feldt, Maria Paasivaara, Daniel Mendez, Stefan Wagner, and Marvin Mu\~{n}oz Bar\'{o}n}
\EventNoEds{5}
\EventLongTitle{20th International Symposium on Empirical Software Engineering and Measurement (ESEM 2026)}
\EventShortTitle{ESEM 2026}
\EventAcronym{ESEM}
\EventYear{2026}
\EventDate{October 8--9, 2026}
\EventLocation{Munich, Germany}
\EventLogo{}
\SeriesVolume{394}
\ArticleNo{22}

\pdfoutput=1
\usepackage[utf8]{inputenc}

\usepackage[T1]{fontenc}
\usepackage{url}
\usepackage{flushend}
\usepackage{xspace}
\usepackage[skip=5pt]{caption}
\usepackage[inline]{enumitem}
\usepackage{ifthen}
\usepackage{balance}
\usepackage{hyperref}
\usepackage{flushend}
\usepackage{colortbl}
\usepackage{listings}

\definecolor{codegreen}{rgb}{0,0.6,0}
\definecolor{codegray}{rgb}{0.5,0.5,0.5}
\definecolor{codepurple}{rgb}{0.58,0,0.82}
\definecolor{backcolour}{rgb}{0.95,0.95,0.92}

\lstdefinestyle{mystyle}{
    commentstyle=\color{codegreen},
    keywordstyle=\color{magenta},
    numberstyle=\tiny\color{codegray},
    stringstyle=\color{codepurple},
    basicstyle=\ttfamily\footnotesize,
    breakatwhitespace=false,         
    breaklines=true,                 
    captionpos=b,                    
    keepspaces=true,                 
    numbers=left,                    
    numbersep=5pt,                  
    showspaces=false,                
    showstringspaces=false,
    showtabs=false,                  
    tabsize=2
}

\usepackage{microtype}
\usepackage{multirow}
\usepackage[]{siunitx}
\newcommand*\np[2][z]{
\ifx z#1%
$\num{#2}$%
\else%
$#2\ #1$
\fi\xspace%
}
\usepackage{fp}
\usepackage{xfp}

\usepackage{graphicx}
\usepackage{subcaption}
\usepackage{pifont}

\usepackage{booktabs}
\usepackage{multirow}
\usepackage{threeparttable}
\usepackage{xcolor}
\usepackage{array,makecell}
\usepackage{diagbox}
\usepackage{tabularx}

\usepackage{mathtools}
\usepackage{amsmath}

\usepackage{tcolorbox}
\usepackage{pgfplots}
\usepackage{pgfplotstable}

\usepackage{enumitem}
\usepackage[framemethod=TikZ]{mdframed}
\usepackage[scaled]{beramono}

\usepgfplotslibrary{statistics}
\usetikzlibrary{calc,patterns,fpu} 
\pgfplotsset{compat=1.14} 

\pgfplotsset{
    /pgf/declare function={
        Floor(\x) = round(\x-0.49);
    },
    show sum on top/.style={
        /pgfplots/scatter/@post marker code/.append code={%
            \path let \p1=($(normalized axis cs:%
                        \pgfkeysvalueof{/data point/x},%
                        \pgfkeysvalueof{/data point/y})%
                        -(normalized axis cs:\pgfkeysvalueof{/data point/x},0)$)
            in node[
                at={(normalized axis cs:%
                        \pgfkeysvalueof{/data point/x},%
                        \pgfkeysvalueof{/data point/y})%
                },
                anchor={-90*sign(\y1)},yshift={sign(\y1)*2pt}
            ]
            {\pgfmathprintnumber[fixed, precision=1]{\pgfkeysvalueof{/data point/y}}};
        },
    }
}

\definecolor{blau_1a}{RGB}{93,133,195}
\definecolor{blau_2a}{RGB}{0,156,218}
\definecolor{gruen_3a}{RGB}{80,182,149}
\definecolor{gruen_4a}{RGB}{175,204,80}
\definecolor{gruen_5a}{RGB}{221,223,72}
\definecolor{orange_6a}{RGB}{255,224,92}
\definecolor{orange_7a}{RGB}{248,186,60}
\definecolor{rot_8a}{RGB}{238,122,52}
\definecolor{rot_9a}{RGB}{233,80,62}
\definecolor{lila_10a}{RGB}{201,48,142}
\definecolor{lila_11a}{RGB}{128,69,151}

\definecolor{blau_1b}{RGB}{0,90,169}
\definecolor{blau_2b}{RGB}{0,131,204}
\definecolor{gruen_3b}{RGB}{0,157,129}
\definecolor{gruen_4b}{RGB}{153,192,0}
\definecolor{gruen_5b}{RGB}{201,212,0}
\definecolor{orange_6b}{RGB}{253,202,0}
\definecolor{orange_7b}{RGB}{245,163,0}
\definecolor{rot_8b}{RGB}{236,101,0}
\definecolor{rot_9b}{RGB}{230,0,26}
\definecolor{lila_10b}{RGB}{166,0,132}
\definecolor{lila_11b}{RGB}{114,16,133}

\newboolean{showcomments}
\setboolean{showcomments}{true}

\newcommand{\ShowAbsoluteNumber}[1]{%
\ifnum #1<10%
{\hspace*{0pt}#1}%
\else%
\ifnum #1<100%
{\hspace*{0pt}#1}%
\else%
\ifnum #1<1000%
{\hspace*{0pt}#1}%
\else%
{\numprint{#1}}%
\fi%
\fi%
\fi%
}

\newcommand{\ShowPercentage}[2]{%
\FPeval\percentage{round(#1/#2*100,0)}%
\FPeval\percentageOneDecimal{round(#1/#2*100,1)}%
\ifnum \percentage=0%
{\np[\%]{\FPprint{percentageOneDecimal}}}%
\else%
\ifnum \percentage<10%
{\np[\%]{\FPprint{percentageOneDecimal}}}%
\else%
{\np[\%]{\FPprint{percentageOneDecimal}}}%
\fi%
\fi%
\xspace
}
\newlength\BARSIZE  
\newcommand{\inlinechart}[2]{%
\FPeval{\BLACKBARSIZE}{#1/#2}\textcolor{black!80}{\rule{\BLACKBARSIZE\BARSIZE}{1.6ex}}%
\FPeval{\BLACKBARSIZE}{1 - (#1/#2)}\textcolor{black!10}{\rule{\BLACKBARSIZE\BARSIZE}{1.6ex}}%
}

\newcommand*\percent[3][v]{%
\ifx q#1%
    \np{#2}/\np{#3}(\ShowPercentage{#2}{#3})\else%
\ifx s#1%
    \ShowPercentage{#2}{#3}\else%
\ifx p#1%
    \np{#2}(\ShowPercentage{#2}{#3})\else%
\ifx c#1%
    \inlinechart{#2}{#3}%
\else%
    \np{#2}%
    \ifx r#1%
        /\np{#3}%
    \fi%
    \hspace*{0.5ex}(\ShowPercentage{#2}{#3}) %
    \inlinechart{#2}{#3}%
    \xspace
\fi\fi\fi\fi%
}

\def\morph{\textsc{Morph}}

\definecolor{mygray}{RGB}{240,240,240}

\newcommand{\answer}[2]{\vspace{.2cm}{\centering\setlength{\fboxrule}{0.1pt}\fbox{\colorbox{mygray}{\parbox{0.95\columnwidth}{$\textbf{Answer to RQ}_{#1}$. #2}}}\vspace{.2cm}}}

\ifthenelse{\boolean{showcomments}}
 { }
        { }

\definecolor{eminence}{RGB}{108,48,130}
\definecolor{weborange}{RGB}{255,165,0}
\definecolor{frenchplum}{RGB}{129,20,82}
\definecolor{darkgreen}{RGB}{10, 92, 10}

\mdfdefinestyle{mpdframe}{
    nobreak                     =true,
    backgroundcolor             =black!3,
    frametitlebackgroundcolor   =black!15,
    frametitlerule              =false,
    roundcorner                 =5pt,
    innermargin                 =0.3cm,
    innerleftmargin             =0.3cm,
    innerrightmargin            =0.3cm,
    innertopmargin              =0.3cm,
    innerbottommargin           =0.3cm,
    shadowsize                  =1pt
}

\begin{document}

\maketitle

\begin{abstract}
\textbf{Background:} Large Language Models (LLMs) are increasingly being applied to Software Engineering (SE) tasks, achieving high accuracy across problems such as clone detection, vulnerability prediction, and code summarization. However, their high computational demands and energy consumption raise sustainability concerns and hinder their use on consumer hardware and resource-constrained platforms. A common way to report the computational cost of an LLM in the literature and industry is to use the number of Floating Point Operations (FLOPs) required to perform a pass over the network.
\textbf{Aims:} This paper investigates the implications of energy-aware knowledge distillation for SE, aiming to improve model efficiency while maintaining performance and to determine whether FLOPs is a reliable energy-aware metric.
\textbf{Method:} We conduct a controlled experiment using \textsc{Morph}, a Many-Objective Optimization-based distillation methodology, to empirically examine whether FLOPs accurately reflect energy consumption in Clone Detection and Vulnerability Prediction tasks. We extend this methodology to include energy-surrogate models that directly estimate CPU and GPU energy consumption during optimization, and we apply \textsc{Morph} to generative tasks using CodeT5+ for code summarization.
\textbf{Results:} Our results show that FLOPs is not always a reliable indicator of energy consumption, and better results can be achieved by using energy-surrogate models. Distilled student models can reduce inference energy consumption by up to 90\% and memory usage by 86\%, with only modest accuracy trade-offs.
\textbf{Conclusions:} Energy-aware knowledge distillation when guided by direct energy surrogates rather than FLOPs can improve the energy consumption, sustainability, and deployability of LLMs for SE applications, enabling efficient models on consumer hardware.
\end{abstract}

\section{Introduction}
Large language models (LLMs) have recently emerged as a highly useful tool for tasks that require language understanding and generation.
Among them, LLMs are capable of supporting an extensive number of Software Engineering (SE) tasks~\cite{fan2023large}, including both classification (e.g., clone detection \cite{10556419}, vulnerability prediction \cite{10.1145/3639476.3639762}) and generation tasks (e.g., code summarization \cite{DBLP:conf/acl/AhmadCRC20} or code completion \cite{10.5555/3618408.3618894, 10.1145/3324884.3416591}).
Their capabilities come from massive scales: billions of parameters trained on terabytes of open-source code from platforms such as GitHub~\cite{lozhkov2024starcoder}.
This scale poses significant computational challenges, including the need for large datacenters with dedicated hardware for parallel computation, such as GPUs \cite{DBLP:conf/mlsys/WuRGAAMCBHBGGOM22}.
Powering these datacenters requires massive amounts of energy: the International Energy Agency projects that datacenters' energy consumption will double over the next 5 years \cite{IEA2025EnergyAI}, further straining existing electrical infrastructure and increasing greenhouse gas emissions and environmental costs~\cite{10.1145/3381831}.


Optimizing LLM energy consumption requires addressing where and how operational costs accumulate. 
Because inference dominates lifetime energy usage for high-volume models~\cite{flopsToEnergy}, relying on large, general-purpose models over long sequences incurs substantial computational and energetic overhead~\cite{10.1145/3630106.3658542}. 
To mitigate this, model compression reduces the compute and memory footprint per query, directly lowering operational energy demands. For SE workflows, such compressed models provide a dual benefit: they minimize runtime power consumption and enable lightweight, local deployment directly inside IDEs, avoiding energy-intensive calls to remote cloud servers~\cite{11029766, 9463109}.

A promising technique that can generate smaller, more efficient models addressing these three aspects is knowledge distillation (KD) \cite{hinton2015distillingknowledgeneuralnetwork}, where a smaller model (\textit{student}) learns the implicit knowledge of a larger model (\textit{teacher}) for a downstream task. By learning to mimic the teacher's output probability distribution, the student model maintains similar accuracy while using considerably fewer resources.
%
In the context of SE tasks, Panichella \cite{11029766} proposes \morph{}, a method that compresses BERT models to smaller sizes using evolutionary algorithms and predictive models (or surrogates) to improve accuracy and metamorphic robustness.
This method achieves state-of-the-art performance on clone detection and vulnerability prediction tasks, surpassing prior work on this topic \cite{10.1145/3639475.3640097}.

While these results are promising, two aspects limit their generalizability. First, \morph{} relies on Floating Point Operations (FLOPs) as a proxy for energy consumption. The approach aims to minimize the FLOPs used by a model to reduce energy requirements. FLOPs measure the number of floating-point operations required for a training or inference pass and can be derived directly from a model’s architecture and hyperparameters. Their simplicity and hardware-agnostic nature make FLOPs attractive, and they have therefore been widely used in the Machine Learning literature as a stand-in for energy consumption in LLM research~\cite{jiao-etal-2020-tinybert, liu-etal-2020-fastbert, li-etal-2021-accelerating}.

However, the validity of FLOPs as a proxy for energy consumption remains debated in the literature, where their correlation to energy usage is heavily dependent on the context. Most FLOPs estimates---such as those from common profiling tools\footnote{\url{https://docs.pytorch.org/docs/stable/profiler}}---capture only the theoretical count of mathematical operations (e.g., matrix multiplications, additions) while ignoring runtime factors that strongly influence energy usage, including memory access patterns, data movement overheads, and hardware-level optimizations on GPUs or TPUs~\cite{flopsToEnergy, flopsCriticism, 10371661}. As a result, FLOPs may not reliably reflect actual energy consumption.


Second, the evaluation for \morph{} is limited to BERT-like models compressed from up to 500 MB to 3 MB. Such size reductions do not necessarily yield proportional energy savings, as energy is subject to diminishing returns due to baseline power consumption or memory bandwidth saturation. Furthermore, BERT is now relatively dated, while newer architectures such as CodeT5~\cite{DBLP:conf/emnlp/WangLGB0H23} and Starcoder 2~\cite{lozhkov2024starcoder} have since emerged, raising questions about generalizability. Moreover, the experiments focus only on two binary classification tasks. More complex tasks involving generation not only require students to capture deeper semantic knowledge but also involve multiple forward passes during autoregressive decoding. This substantially increases inference costs and makes energy estimation more challenging.

To address these limitations, we first revisit the efficiency assumptions behind \morph{} and then extend the method to optimize for measured energy directly and to handle generation tasks. Concretely, we ask the following research questions:

\textbf{$\textbf{RQ}_{1}$}: \textit{Are FLOPs a good proxy of energy consumption in the context of model distillation using Many Objective Optimization?}

To answer this question, we replicate the experiments from the original paper and measure energy during the training and inference phases for different student models with varying FLOPs, using an energy profiler to capture CPU and GPU energy. Statistical testing shows that FLOPs do not correlate well with energy measurements from the profiler, and that other factors, such as specific hyperparameters, may provide stronger correlations with energy.

Based on these results, we modify \morph{} to improve its energy estimation. Instead of relying on FLOPs, we introduce measured energy (from the profiler) as a new optimization objective and build a surrogate model to predict energy consumption from hyperparameter combinations. This naturally motivates our next question:

\textbf{$\textbf{RQ}_{2}$}: \textit{How energy efficient are \morph{} models when using actual energy as an objective?}

This lets \morph{} search for configurations that are efficient with respect to observed hardware behavior rather than architectural estimates alone.

Finally, we examine whether Many-Objective Optimized Distillation can be effective beyond classification, in more complex generative tasks that may require deeper semantic knowledge. To investigate this, we extend the \morph{} methodology with a new distillation pipeline and investigate:

\textbf{$\textbf{RQ}_{3}$}: \textit{How effective is \morph{} at reducing LLMs costs for generation tasks, and what is the accuracy tradeoff?}

We evaluate the modified method with CodeT5+~\cite{DBLP:conf/emnlp/WangLGB0H23}, distilling models for code summarization with energy as an additional optimization objective. We compare the resulting students with the teacher model in terms of accuracy (ROUGE-L), size, and energy consumption.

In summary, this paper makes the following contributions:
\begin{itemize}
    \item Empirical study of FLOPs vs. energy. We replicate \morph{}’s experiments and show that FLOPs correlate inconsistently with actual energy consumption in SE tasks.
    \item Energy-aware extension of \morph{}. We introduce measured energy as an optimization objective, with surrogate models to estimate energy from hyperparameters.
    \item Application of the extended method to CodeT5+ for code summarization, achieving up to 90\% lower energy and 86\% smaller memory footprint, with modest accuracy trade-offs.
\end{itemize}

\section{Background\label{sec:back}}

Knowledge distillation \cite{hinton2015distillingknowledgeneuralnetwork} is a technique for compressing large pretrained models into smaller, more efficient models, ideally with minimal impact on accuracy.
The smaller model serves as the student, and the larger model acts as the teacher.
Instead of learning directly from the dataset, the student model learns to predict responses similar to the teacher's.
To do this, the training dataset is fed into the teacher and the student, and the student's loss is computed to minimize the distance between the student's and teacher's output probability distribution.
Distillation can be done with generic pretraining tasks, to compress general-purpose models~\cite{deepseekai2025deepseekr1incentivizingreasoningcapability}, or as task-specific distillation, where the student learns from a teacher finetuned to perform a specific task, and can be used for the same task without any additional finetuning.

Knowledge distillation brings multiple advantages. The dataset size required for distilling a student model is smaller, as is the number of epochs needed to reach convergence, which lowers training costs.
However, it requires an already pretrained, high-performing teacher.
Additionally, for task-specific distillation, depending on the task's complexity, some pretrained weights are required for the student model.
For example, in binary classification tasks, the student model's weights can be randomly initialized, whereas more complex tasks, such as generation, may require pretraining or copying weights from the teacher.

On top of this, finding the optimal hyperparameter set for the student model can be challenging. 
The possible combination of hyperparameters, such as the number of layers or attention heads, makes the search space extremely large.
To tackle this, several approaches model this search as an optimization problem with one or more objectives, and the different hyperparameters as variables.
The state-of-the-art implementation \morph{} \cite{11029766} employs a Many-Objective optimization approach to compress models for two binary classification SE tasks: Clone Detection and Vulnerability Prediction.
It defines 4 objectives for the student model: maximum accuracy, minimum memory footprint, minimum FLOPs, and maximum robustness.
For robustness, the paper uses metamorphic testing, which involves creating new data samples by applying transformations to the input data that preserve the code's functionality and semantics, such as replacing English words with synonyms or expanding acronyms.
Robustness is measured using prediction flips: given a code sample modified by metamorphic transformations, prediction flips determine whether the model maintains the original prediction.

To solve the Many-objective optimization problem, \morph{} employs AGE-MOEA \cite{10.1145/3321707.3321839}, an evolutionary algorithm that evolves an initial population of solutions and converges towards the Pareto front.
In Many-objective optimization, the Pareto front is the set of solutions that achieve optimal trade-offs among the objectives. These are solutions in which improving one objective would negatively impact the others.
To explore the population space without distilling and evaluating every possible student, \morph{} employs surrogate models to improve accuracy and reduce prediction flips. 
By uniformly sampling the configuration space and training a set of student models, it builds two surrogate models to predict the accuracy and prediction flips of a given configuration, without training that configuration

\section{Related Work}

\textbf{Many-Objective Knowledge Distillation for LLMs\label{sec:rel-morph}}.
Multi- and Many-Objective optimization for Knowledge Distillation is a relatively recent technique that explores the configuration space of student models to optimize for two or more objectives, including maximizing accuracy and minimizing computational cost and size.
The first proponent was Shi et al. \cite{10.1145/3551349.3556964} with \textsc{Compressor}, which used a single-objective genetic algorithm to distill binary classification models, minimizing a combination of Giga FLOPs and model size.
Later, they introduced \textsc{Avatar} \cite{10.1145/3639475.3640097}, which uses a multi-objective genetic algorithm with 3 objectives: minimize FLOPs and model size, and maximize accuracy.

\morph{} \cite{11029766} builds on top of this previous work, introducing robustness as an additional objective, which is done using metamorphic testing, as explained in Section \ref{sec:back}. With its additions --- metamorphic testing, surrogate models, and Many-Objective optimization --- \morph{} achieves state-of-the-art performance compared to \textsc{Avatar}, but leaves some questions unanswered.
First, it uses FLOPs as an optimization objective for energy and sustainability, as does the previous work.
However, the relationship between FLOPs and energy consumption is not straightforward and depends on factors such as the target hardware and the potential optimizations it enables.
In this work, we aim to determine whether FLOPs is an adequate objective to minimize when reducing the energy consumption of a distilled model.

Another topic these approaches do not address is their validity for generation tasks. 
Previous work only performs distillation for binary classification tasks, namely, Clone Detection and Vulnerability Prediction.
These tasks are much simpler than generation tasks and require less complex models, allowing for more aggressive compression during distillation.
In this paper, we will apply this technique to a generation task, Code Summarization, and investigate whether the advantages it offers can be applied to generation tasks.
\smallskip

\textbf{FLOPs as proxy for energy consumption}.
The validity of floating-point operations (FLOPs) as a proxy for an AI model's energy consumption is an ongoing debate in the field.
FLOPs were first proposed as an energy proxy in the early stages of Green AI \cite{10.1145/3381831, DBLP:journals/corr/abs-1910-09700} because they can be easily calculated solely from architectural details.
Desislavov et al. \cite{flopsToEnergy} perform a study on inference energy consumption for a set of AI models for Computer Vision and Natural Language Processing tasks.
Here, the authors propose using FLOPs as a proxy for energy consumption, relating the FLOPs used by a model during inference to hardware FLOPs and combining them with GPU hardware specifications, specifically TFLOP/s and power.
However, this study fails to address the gap between logical and hardware FLOPs, as it does not account for potential hardware optimizations enabled by GPUs or TPUs.

Asperti et al. \cite{flopsCriticism} investigated the validity of FLOPs as a proxy metric for energy consumption of Convolutional Neural Networks (CNN). 
They show that FLOPs do not correlate well with energy consumption and propose a metric, $\alpha$-\text{FLOPS}, that accounts for hardware and model architecture to estimate energy more accurately.
However, this work focuses solely on CNNs and does not consider other architectures, such as transformers, widely used by LLMs.
A similar study by del Rey et al. \cite{10371661}
on multiple Deep Neural Network architectures shows that FLOPs might be a better estimator of GPU usage than energy consumption.

A gap remains regarding whether the correlation between FLOPs and energy consumption observed in previous studies is still applicable to LLMs. 
Despite this, many LLM-related papers still use it to reflect energy costs of models, as seen in Section \ref{sec:rel-morph} with \morph{}, but also many others \cite{jiao-etal-2020-tinybert, liu-etal-2020-fastbert, li-etal-2021-accelerating}.
This paper aims to shed light on this topic and investigate whether FLOPs are a viable proxy for energy consumption in certain LLM tasks.



%



\section{Methodology\label{sec:methodology}}


This section introduces the changes applied to the original \morph{} methodology to investigate our research questions.
For $\textbf{RQ}_{1}$, we maintain the original approach, and empirically measure energy usage of the distilled models in the Pareto front.
For $\textbf{RQ}_{2}$, we extend this approach by adding energy as an additional objective in the function.
For $\textbf{RQ}_{3}$, we take this one step further and use our modified approach for a more complex generation task.

\subsection{Adding Energy as Optimization Objective \label{sec:energy_objective}}
For research question $\textbf{RQ}_{2}$, we closely maintain the original methodology steps. The only change we introduce is to redefine the efficiency function, using energy rather than FLOPs to measure a model's energy efficiency.

To do this, we adapt the surrogate model methodology proposed in the original paper for predicting accuracy and the number of prediction flips.
We sample a representative subset of the hyperparameter subset using Latin Hypercube Sampling (LHS), which provides a uniformly distributed set of configurations across the sampling space.
For each hyperparameter set, we assess median CPU and GPU energy consumption across multiple runs, following the protocol described in \autoref{sec:energy_protocol}, along with accuracy and prediction flips.
Then, we use the energy measurements to train two additional regression models to predict GPU and CPU energy consumption for a given set of hyperparameters.

During the optimization phase, instead of computing and minimizing FLOPs for the sampled population, we predict CPU and GPU energy consumption and use the sum as our new minimization objective.
If the surrogate models are sufficiently accurate, this serves as an alternative proxy metric for energy.


\subsection{Morph for Generation}
To investigate $\textbf{RQ}_3$, we adapt \morph{}'s objective function to work for a generation task. 
Concretely, we will test \morph{} and CodeT5+ performance for the Code Summarization task.
To evaluate the efficacy of a student model for this task, we want the student model to produce summaries as close as possible to those in the dataset, while minimizing energy usage and model size.
Therefore, we redefine the problem and objective function as follows:

\textit{Definition: Given a LLM teacher model, find a hyperparameter configuration} $C = [c_1,\dots,c_m]$ \textit{for the student model} $S$ \textit{that optimizes the following objectives:}
\begin{equation*}
\begin{cases} 
min \ O_1 = & size(S) \\
min \ O_2 = & efficiency(S) \\
max \ O_3 = & effectiveness(S,V)
\end{cases}
\vspace{-2mm}
\end{equation*}
where $V$ is the validation set.
The main changes to the original formulation lie in the lack of a robustness metric and the measurement of efficiency and effectiveness.
To measure effectiveness, we require metrics that measure similarity between generated texts and ground truth.
Ideally, we would compare the summaries' semantic similarity using human judgment or an LLM serving as a judge. However, since the methodology requires training and testing a relatively high number of models, this would significantly increase the time and resources needed.
Therefore, we select an ngram-based metric.
To partially mitigate the lack of human judgment, we select ROUGE-L as our accuracy metric. 
ROUGE-L has been shown to align better with human judgment \cite{EVTIKHIEV2023111741} than other state-of-the-art metrics, such as BLEU, METEOR, or CodeBLEU, for code-related tasks.

To assess robustness, the original paper used the number of prediction flips as a metric, since it dealt with binary classification tasks.
In the case of summarization, robustness is much more difficult to assess, since there is not a single correct summary for a piece of code. 
Furthermore, introducing metamorphic code changes, such as synonymizing parameter names, can yield the same changes in the summary, maintaining semantic similarity while using different words.
This could also potentially be addressed with human or LLM judgment, but such techniques do not fit the size of these methodologies.

Finally, to measure efficiency, we will use energy measurements of the inference process in Joules, rather than FLOPs, following the process defined in the previous section.
Using the same inference dataset, we measure the energy required for each student to complete it.

To create the surrogate models, we follow the technique described in Section \ref{sec:energy_objective} for the first two research questions and GraphCodeBERT. 
We sample the hyperparameter space using LHS, distill the models, and measure their ROUGE-L scores on the validation set and their inference energy consumption.

\textbf{Weight subcloning}. Because clone detection and vulnerability prediction are binary classification tasks, the student model weights were randomly initialized, as the number of epochs required for convergence is relatively small.
However, for summarization, the implicit knowledge the student model needs to learn is larger, and convergence takes longer with random initialization. 
Additionally, in our preliminary studies, we observed that randomly initialized students converge to generic responses independently of the input.
To solve this, we clone part of the teacher model's weights into the student model before starting the distillation process.

To copy the most important weights, we apply a weight subcloning technique \cite{DBLP:journals/corr/abs-2312-09299}. 
This technique runs inference on the teacher model over a small subset of the validation set (around 80,000 tokens) and captures the input and output magnitudes of different neurons and attention heads.
Based on these inputs and outputs, it ranks neurons and layers by importance, and copies the most important weights into the student model, respecting the new architecture.
According to the study, student models initialized this way achieve 4x faster training times.

\textbf{Teacher Logit extraction and Distillation Loss}. In the original \morph{} for binary classification tasks, inference is run on the teacher over the training set, and the two logits per sample (positive probability, negative probability) are saved. 
These logits are later used during offline distillation, where only the student model needs to be instantiated in memory.

Compared to binary classification, generative tasks can be viewed as classification over the vocabulary size at each output position, making it infeasible to store logits. 
In our preliminary experiments, the logits for only 64 training samples already required about 1 GB of disk space.
Since it is unfeasible to store all the logits, we perform online distillation.
We instantiate the teacher and student models on the same GPU and, for each batch, first perform inference on the teacher to obtain the logits, then perform the training step on the student.
While this helps avoid storage problems with logits, it increases computational demand, since we need two models instantiated and running inference simultaneously.


To compute the loss between the student output, teacher output, and ground truth, we used the average of the Kullback-Leibler Divergence (KLD) and cross-entropy loss.
KLD is a loss function to compute the difference between two probability distributions. In this case, it takes the student and teacher logits and compares the probability distributions for each output token across the entire vocabulary.
This loss is the standard used across Knowledge Distillation tasks~\cite{kl1,kl2} and it has been shown to work well with LLMs~\cite{klsupport}.

To ensure the student models remain aligned with the ground truth, we include the cross-entropy loss between the student predictions and the ground-truth probability distribution.

\section{Evaluation}

The goal of this study is twofold: (1) to validate the use of \morph{} for optimizing LLMs with respect to energy consumption, and (2) to assess its applicability to generation tasks.
\begin{itemize}
    \item $\textbf{RQ}_{1}$: \textit{Are FLOPs a good proxy of energy consumption in the context of model distillation using Many Objective Optimization?}
    \item $\textbf{RQ}_{2}$: \textit{How energy efficient are \morph{} models when using actual energy as an objective?}
    \item $\textbf{RQ}_{3}$: \textit{How effective is \morph{} at reducing LLMs costs for generation tasks, and what is the accuracy tradeoff?}
\end{itemize}


The objective of $\textbf{RQ}_{1}$ is to empirically assess the validity of FLOPs as a proxy for energy consumption by examining their correlation in models generated by \morph{}, thereby identifying potential weaknesses in the existing distillation approaches for code-specific LLMs~\cite{11029766, 10.1145/3639475.3640097}.
%
$\textbf{RQ}_{2}$ evaluates our proposed modifications to \morph{}, which incorporate energy directly into the optimization. The goal is to determine whether these models are more energy-efficient than those distilled using the standard approach.

Finally, the objective of $\textbf{RQ}_{3}$ is to verify the applicability of our modified \morph{} method to generation tasks for software engineering. 
The original \morph{} approach is only tested for binary classification tasks.
Text generation is a much more complex task, and achieving results similar to those presented in the original paper might not be possible.
With this question, we aim to verify whether we can train LLM models suitable for generation tasks that preserve accuracy while being significantly smaller and more energy-efficient.
\subsection{Downstream tasks}
For $\textbf{RQ}_1$, to keep alignment with the \morph{} experiments \cite{11029766}, we evaluate energy consumption of \morph{} models on the same software engineering tasks and datasets: Clone Detection and Vulnerability Prediction. 
With this, we can replicate previous results exactly and draw valid comparisons.
For the clone detection task, the dataset used is BigCloneBench \cite{6976121}, a collection of Java code clones obtained by mining 25,000 Java projects from SourceForge and Google Code.
For vulnerability prediction, the dataset used is Devign \cite{DBLP:conf/nips/ZhouLSD019}, which contains C code from two open-source libraries, FFmpeg and Qemu.
As explained in previous research, this selection avoids data leakage problems \cite{11029766}.

As a generation task, we chose Code Summarization.
We chose this task because it is quick to evaluate (does not require running test cases, as in code generation) and provides an objective score quickly via ROUGE-L.
However, it is still a relevant task that is used for automated commit messages or pull requests.
We use two separate datasets to avoid data leakage problems.
For finetuning the teacher, we use the Python split of the CodeXGLUE code summarization dataset \cite{DBLP:journals/corr/abs-2102-04664}.
This dataset is extracted from CodeSearchNet and comes from publicly available, open-source GitHub repositories.

However, CodeT5+, the model used for summarization, has been pretrained on CodeSearchNet data, so we cannot use the validation and test splits of CodeXGLUE for evaluation of the model due to data leakage.
To solve this problem, we will extract validation data from The Heap \cite{DBLP:conf/forge/KatzyPDI25}. This is a contamination-free dataset extracted from GitHub projects with non-permissive licenses and has been de-duplicated against the most popular training datasets.
Using a Python AST parser, we extract functions with docstrings and use them as our evaluation dataset.

\subsection{Large Language Models}
We employ two pretrained models for these experiments, GraphCodeBERT \cite{DBLP:journals/corr/abs-2009-08366}, and CodeT5+ \cite{DBLP:conf/emnlp/WangLGB0H23}.
GraphCodeBERT is a bimodal model pretrained on natural-language and programming-language data, based on BERT and CodeBERT \cite{DBLP:journals/corr/abs-2002-08155}, and extends the latter's capabilities by incorporating information from the code snippet's data flow graph.
In the original \morph{} study \cite{11029766}, both CodeBERT and GraphCodeBERT are tested.
However, for $\textbf{RQ}_{1}$, based on the architecture similarity and the study's results, we replicate the results only for GraphCodeBERT, since the conclusions from this should be applicable to both models.

For the generation tasks, we use CodeT5+ \cite{DBLP:conf/emnlp/WangLGB0H23}. This model is based on the Google T5 model and has been pretrained on multiple code-related tasks, including code summarization, using the CodeSearchNet and BigCode datasets. 
We chose this model because it is more recent than GraphCodeBERT and offers better capabilities.
This family also provides small versions, from which we choose the model with 220M parameters. 
This choice is motivated by our hardware constraints and the need to perform online distillation with 2 models loaded simultaneously. 
Our experiments also require multiple distillations to create surrogate models and multiple energy measurement runs. Using a smaller model reduces the execution times of the experimental pipeline.

\subsection{Empirical Methodology\label{sec:emp}}
To answer $\textbf{RQ}_{1}$, we sample models with varying FLOPs, measure their energy consumption, and analyze the correlation between the two. To obtain these models, we generate a Pareto front of optimal solutions using the original \morph{} approach. In many-objective optimization, the Pareto front consists of solutions for which no objective can be improved without sacrificing at least one other objective.
Therefore, sampling the Pareto front will yield a diverse set of models with varying FLOP requirements. After measuring the energy consumption of these models, we use statistical methods, such as ANOVA, to identify correlations among energy, FLOPs, and other hyperparameters.


To answer $\textbf{RQ}_{2}$, we compare the models generated by the original \morph{} methodology with a modified method that substitutes the FLOPs objective with actual energy consumption predictions, using the surrogate model to predict energy consumption in inference, similar to the surrogate method used for prediction flips.
We compare both methodologies using the metrics from the original paper (\textit{size}, \textit{accuracy}, \textit{robustness}, and \textit{FLOPs}) as well as \textit{inference energy consumption} to determine whether models actually consume less energy during inference.
We run the original \morph{} and the modified version, generating new Pareto fronts and models for both cases.
Given the random nature of the MOP optimization algorithm, we distill 20 models and report the median results for the previous metrics, along with the interquartile range.
To compare with the original \morph{} results, we use the same model selection criteria, choosing the model with size closest to 3MB.
We analyze the results using the Wilcoxon signed-rank test \cite{wilcoxon} to determine whether there is a statistically significant difference in energy consumption between standard and modified \morph{}, and the Vargha-Delaney $\hat{A}_{12}$ statistic~\cite{a12} to determine the effect size.









Finally, to answer $\textbf{RQ}_{3}$, we distill a CodeT5+ model using \morph{} and the energy objective, and compare the student model with the teacher model.
We compare based on three metrics: accuracy (using ROUGE-L), size, and energy efficiency.
We also manually inspect some test samples to evaluate the models' performance based on human judgment.

\subsection{Energy Measurement Setup\label{sec:energy_protocol}}



For our experiments, we measure only energy usage during the inference step. We exclude tokenization and decoding since they depend on the tokenizer, which is not affected by the distillation.
Measuring the energy consumption of an AI model is difficult because software cannot run in a vacuum; it depends on its environment: operating system, dependencies, memory, and hardware.
Additionally, the hardware can be affected by external constraints, such as room temperature. Consequently, obtaining a precise measurement with a single execution is not guaranteed.

We follow existing guidelines to mitigate external factors~\cite{energyMeasurementsGuide}. The first is to make execution times long enough to reduce variance in energy measurements. Accordingly, we assess inference energy of the distilled models using an automated script that: (1) loads the model and test dataset, and (2) runs inference on the dataset multiple times. The multiple inferences are run so the energy costs of loading the model do not overshadow the inference costs.
Another mitigation is to repeat measurements for each model in a randomized order, with short pauses in between, to reduce variation caused by unexpected system activity. Following the guidelines, we perform 20 executions~\cite{energyMeasurementsGuide} and report the median energy consumption across runs.
Between measurements, we reset the GPU context to ensure no cached results affect the measurements.

The experiments are run on a server equipped with an AMD Ryzen 9 7900X processor, 64 GB of RAM, and an NVIDIA GeForce RTX 4090 GPU. The server runs Ubuntu 22.04.3, with Linux kernel version 6.2.0, and Docker 24.0.5.
To measure CPU and GPU energy consumption, we use EnergiBridge \cite{sallou2023energibridge}, a software profiling tool compatible with AMD CPUs and Nvidia GPUs, among others, that uses RAPL and NVML under the hood.

\subsection{Parameter settings}
For the components of the methodology that remain unchanged from the original implementation, we maintain the parameter configuration from the original paper \cite{11029766}. This includes the parameters for population creation and mutation, optimization algorithm, and hyperparameter space for GraphCodeBERT.
We also replicate the configuration of the original paper for the new energy surrogate models, using Gradient Boosting Regression, and grid search to optimize GBR hyperparameters accross the following predefined parameter space: (1) number of decision trees $\in \{50, 100, 150\}$; (2) learning rate $\in \{0.05, 0.10, 0.15, 0.20, 0.25\}$; (3) maximum depth of each decision 3 $\in \{2,3,4,5\}$; (4) the minimum number of samples in internal node $\in \{2,3,4\}$; (5) minimum number of samples in leaf nodes $\in \{1,2,3\}$.

For Code Summarization distillation in CodeT5+ we use the following configuration space: (1) number of hidden layers $\in [1,12]$; (2) hidden activation $\in \{gelu, relu, silu, \textit{gelu-new}\}$; (3) number of decoder layers $\in [1,12]$; (4) hidden size $\in [16,768]$; (5) number of attention heads $\in [1,12]$; (6) projection size $\in [1,64]$; (7) intermediate size $\in [16,3072]$; (8) relative attention buckets $\in [4, 32]$; (9) relative attention max distance $\in [32, 128]$; (10) dropout rate $\in \{0.1,0.2,0.3\}$; (11) feed forward projection $\in \{relu, \textit{gated-relu}\}$; (12) learning rate $\in \{0.001,0.0001,0.00005\}$; (13) batch size $\in \{8,16\}$.
This configuration's maximum values are equivalent in size to the teacher model (CodeT5+ 220M).
This way, student models will be no larger than the teacher model.
We chose this because, a priori, we do not know the level of compression that can be achieved for this task and model, and it may be that the maximum accuracy lies in the teacher's model size.

Additionally, compared to the original paper, we removed hyperparameters for the tokenizer, since modifying it would require retraining the teacher model, since the number of logits equals the vocabulary size for each token.

Finally, for inference during evaluation, we use beam-search multinomial sampling with 5 beams.
\section{Results}
\subsection{RQ1: Correlation of FLOPs and Energy Consumption}

\begin{figure*}[t]
    \centering
    \begin{subfigure}[b]{0.49\textwidth}
        \centering
        \includegraphics[width=\textwidth]{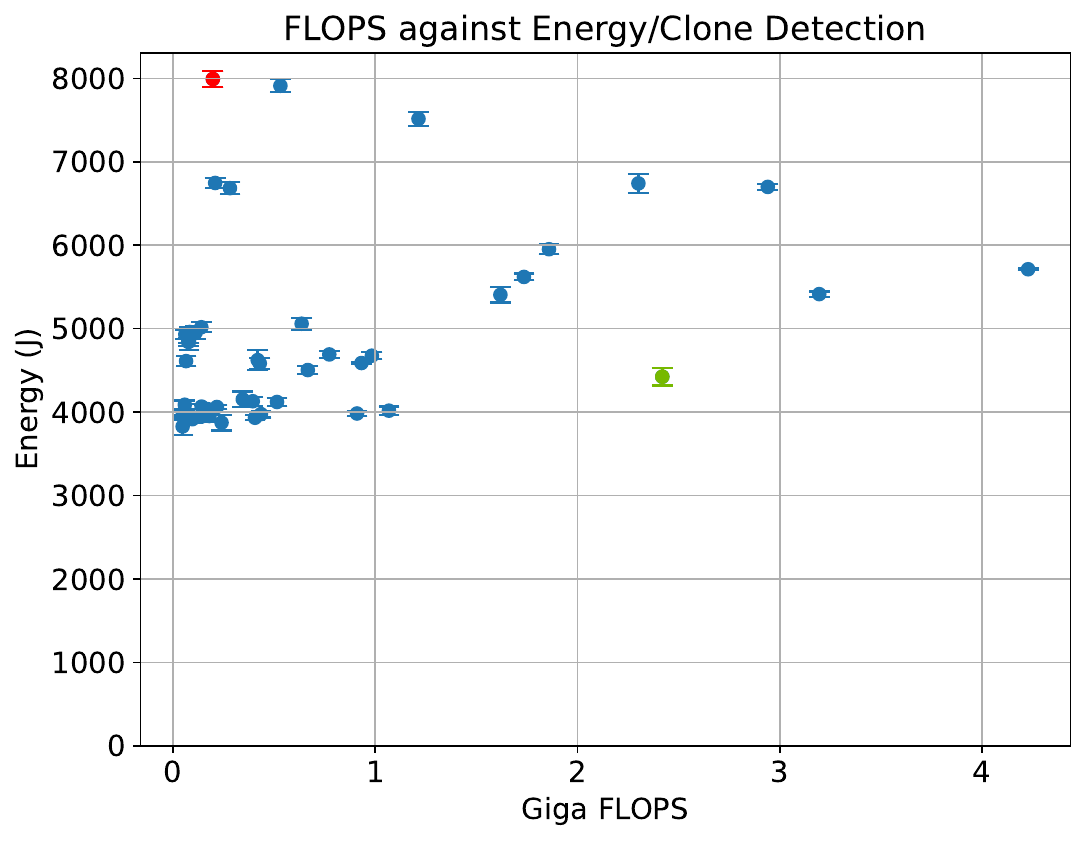}
        \caption{Clone Detection}
        \label{fig:CD_FLOPS_Energy}
    \end{subfigure}
    \hfill
    \begin{subfigure}[b]{0.49\textwidth}
        \centering
        \includegraphics[width=\textwidth]{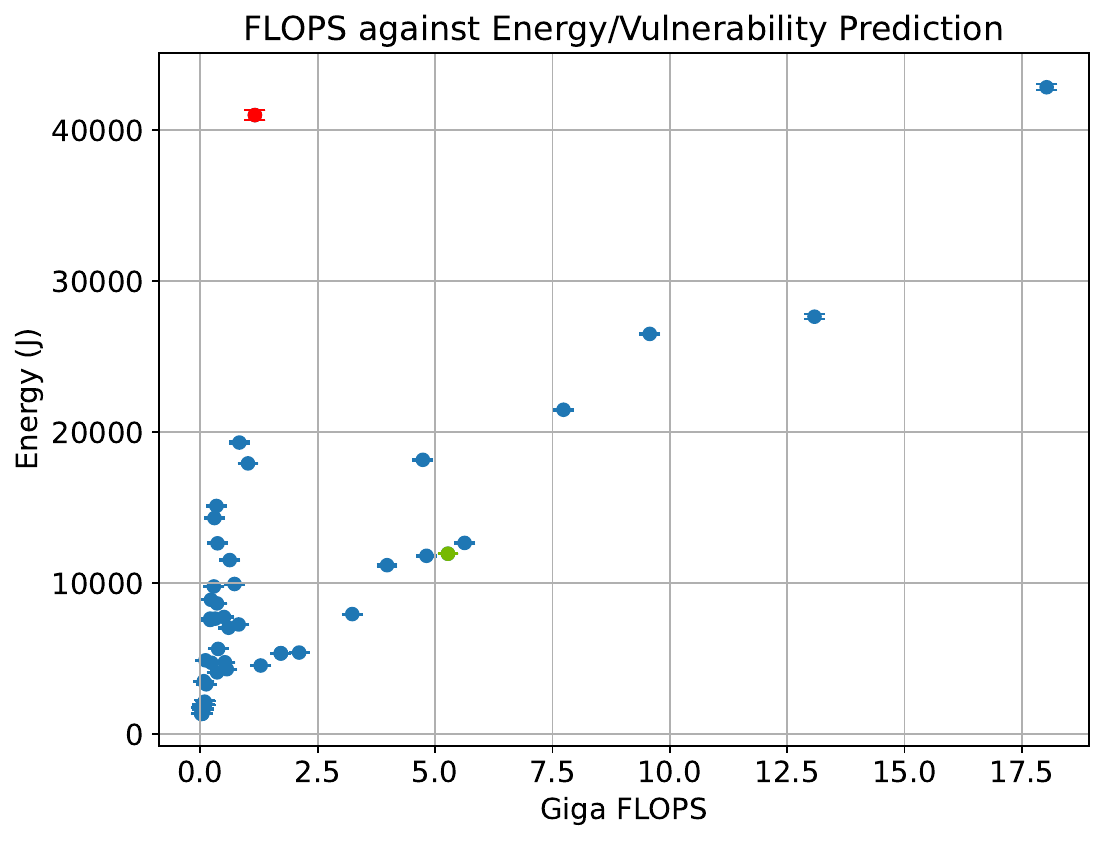}
        \caption{Vulnerability Prediction}
        \label{fig:VD_FLOPS_Energy}
    \end{subfigure}
    \caption{Inference energy usage against FLOPs}
    \label{fig:FLOPS_Energy_Pareto}
\end{figure*}

We ran \morph{} with its default configuration and extracted details of the Pareto front for the Clone Detection and Vulnerability Prediction tasks.
For each Pareto front, we ran energy tests on inference, using the same test set for every model and testing each model multiple times in a randomized order, as explained in Section \ref{sec:energy_protocol}.
Figure \ref{fig:FLOPS_Energy_Pareto} shows a scatterplot of total inference energy consumption vs FLOPs for the Pareto Front of the Clone Detection (Figure \ref{fig:CD_FLOPS_Energy}) and Vulnerability Prediction (Figure \ref{fig:VD_FLOPS_Energy}) tasks.

For the Clone Detection task, the figure shows no clear correlation between FLOPs and Energy.
Most models with fewer Giga FLOPs accumulate around the 4,000 J mark, but there are many models that do not show a straight relationship between FLOPs and energy.
For example, the most expensive model uses around 8,001 J while their computational cost is 0.20 Giga FLOPs (highlighted in red), while a model with 2.4 Giga FLOPs uses 4,423 J, around 44 \% less energy (highlighted in green).

For the Vulnerability Prediction task (Figure \ref{fig:VD_FLOPS_Energy}), the results reveal a different behavior and show a much clearer correlation between FLOPs and energy consumption, with a higher number of FLOPs translating into higher inference energy.
However, despite a clearer correlation, there is again no straightforward relationship between FLOPs and energy consumption.
Similarly to Clone Detection, some models use significantly less energy than others with higher FLOP counts.
For example, a model with 1.16 GFLOPs is using 41,122 J (highlighted in red), while a model with 5.27 GFLOPs is using 11,945 J, or 70\% less energy (highlighted in green).
Relying solely on this correlation may obscure outlier models that are far more energy efficient or, conversely, favor models that are substantially less efficient.

Additionally, during our experiments, we tracked and made surrogate models for both CPU and GPU energy. 
However, we observed that CPU power usage is mostly constant, and total energy only depends on inference time.
This means that the bulk of the processing was happening on the GPU.
Higher CPU energy comes only from measuring the same power usage over a longer period.
Therefore, for these results, we report total energy consumption (CPU + GPU), since the CPU's energy consumption is relevant but does not show a clear trend on its own.

\begin{table}[t]
\centering
\caption{ANOVA test results against inference energy for Clone Detection (CD) and Vulnerability Prediction (VP). Grey denotes significance. For CD, all models in the Pareto front have Num Hidden Layers = 1}
\label{tab:anova}
\small
\begin{tabular}{lcc}
\hline
\textbf{Hyperparameter} & \textbf{CD} & \textbf{VP} \\
\hline
Tokenizer               & 0.06  & 0.82 \\
Vocabulary Size              & \cellcolor{lightgray} 0.03 & \cellcolor{lightgray}0.04 \\
Num Hidden Layers       & ---                   & \cellcolor{lightgray}8.26e-11 \\
Hidden Size             & \cellcolor{lightgray}2.95e-5 & \cellcolor{lightgray}1.89e-4 \\
Hidden Activation             & \cellcolor{lightgray}0.01 & \cellcolor{lightgray}2.23e-4 \\
Number of Attention Heads     & \cellcolor{lightgray}3.69e-14 & \cellcolor{lightgray}6.54e-09 \\
Max Sequence Length     & \cellcolor{lightgray}6.97e-08 & \cellcolor{lightgray}1.71e-05 \\
Position Embedding Type & \cellcolor{lightgray}3.71e-3 & 0.11 \\
Learning Rate           & 0.53             & 0.906290 \\
FLOPS                   & 0.16              & \cellcolor{lightgray}6.48e-07 \\
\hline
\end{tabular}
\end{table}

Table \ref{tab:anova} reports the results of the ANOVA test for inference energy against the different hyperparameters of the model, with p-values $ <0.05$ indicating a statistical correlation between inference energy and the hyperparameter.
For Clone Detection, the ANOVA test results  confirm that there is no statistical correlation between FLOPs and energy, with a $p$-value=$0.17$.
Additionally, the results reveal that other hyperparameters, such as Hidden Size, Number of Attention Heads, and Max Sequence Length, show a high correlation with energy ($ p$-value $ \ll 0.05$).
However, despite these hyperparameters being part of the FLOPs computation, their correlation with energy does not transfer to the FLOPs metric.
From the Figure, it is clear that higher FLOPs do not always translate to higher energy consumption.

For the Vulnerability Prediction task, the ANOVA results confirm the correlation observed in the figure, with a $p$-value of $6.48e-07$ for the FLOPs metric against energy consumption. The test also confirms the correlations with the other hyperparameters already observed in Clone Detection.

The results for both tasks suggest that FLOPs can be a somewhat accurate proxy for energy consumption, but other factors that affect energy consumption are not reflected in this metric.
In the original \morph{} methodology, the same FLOPs computation algorithm taken from Clark et al. \cite{DBLP:journals/corr/abs-2003-10555} is used for both classification tasks.
This suggests that FLOPs is not a task-agnostic proxy for energy consumption and fails to capture certain behaviors in the Clone Detection task that strongly influence energy consumption.

\answer{1}{For Clone Detection, FLOPs fail as an energy proxy ($p=0.17$), while for Vulnerability Prediction FLOPs are statistically associated with energy ($p=6.48\times10^{-7}$). This task-dependent contrast shows that FLOPs are not a reliable task-agnostic proxy, and the observed spread of models with similar FLOPs but different energy confirms that FLOPs are not directly proportional to energy consumption.}

\subsection{RQ2: Substituting FLOPS with surrogate energy prediction}

\begin{table*}[ht]
\centering
\caption{Results of Morph using FLOPs and surrogate energy as optimization objectives. The best result for each metric is highlighted in grey.}
\label{tab:results_rq12}
\renewcommand{\arraystretch}{1.2}
\resizebox{\textwidth}{!}{%
\begin{tabular}{l l cc cc cc cc cc}
\toprule
\multirow{2}{*}{\textbf{Task}} & 
\multirow{2}{*}{\begin{tabular}[c]{@{}l@{}}\textbf{Objective}\\\textbf{Metric}\end{tabular}} & 
\multicolumn{2}{c}{\textbf{Accuracy (\%)}} &
\multicolumn{2}{c}{\textbf{Size (in MB)}} &
\multicolumn{2}{c}{\# \textbf{Pred. Flips}} &
\multicolumn{2}{c}{\textbf{Giga FLOPs}} &
\multicolumn{2}{c}{\textbf{Energy (J)}}\\
\cmidrule(lr){3-4} \cmidrule(lr){5-6} \cmidrule(lr){7-8} \cmidrule(lr){9-10} \cmidrule(lr){11-12}
 & & \textbf{Median} & \textbf{Diff.} & \textbf{Median} & \textbf{Diff.} & \textbf{Median} & \textbf{Diff.} & \textbf{Median} & \textbf{Diff.} &
 \textbf{Median} & \textbf{Diff.} \\
\midrule
Clone & FLOPS & 95.26 (2.42) &  & 3.00 (0.01) & & 43.00 (14.25) &  & 0.79 (0.36) &  & 5355.83 (1391) &  \\

Detection & Energy & \cellcolor{lightgray}95.95 (0.97) & +6.09\% & 3.01 (0.06) & +0.01 & \cellcolor{lightgray} 37.00 (27.00) & -57\% & \cellcolor{lightgray}0.98 (0.24) & +21\% & \cellcolor{lightgray}3604.50 (824) & -39\% \\
\cmidrule(l){1-12}

Vulnerability & FLOPS & 55.60 (4.77) & & 2.98 (0.11) &  & \cellcolor{lightgray}143.00 (257.00) &  & \cellcolor{lightgray}0.68 (0.47) &  & \cellcolor{lightgray} 10505.50 (6762.75) & \\

Prediction & Energy & \cellcolor{lightgray} 56.20 (4.50) & -0.27\% & 2.99 (0.13) & +0.01 & 218.00 (260.00) & -33\% & 0.98 (0.20) & +36\% &  14219.00 (8568.75) & +30\%\\
\bottomrule
\end{tabular}%
}
\end{table*}

We apply \morph{} to produce 20 student models optimized for FLOPs and energy, respectively. Due to the method's stochastic nature, each model yields different trade-offs. Table~\ref{tab:results_rq12} reports median and interquartile range for accuracy, model size, prediction flips, GFLOPs, and energy. As per Section~\ref{sec:emp}, we select the model with size closest to 3MB.

For accuracy and prediction flips, neither optimization strategy significantly outperforms the other. Energy optimization yields slightly higher accuracy, but the difference is not statistically significant ($p$-value$>$0.3) for both Clone Detection and Vulnerability Prediction. Conversely, FLOPs optimization shows fewer prediction flips, but again the difference is not significant ($p$-value$>$0.5).

An opposite but similar effect occurs with robustness: the original FLOPs optimization outperforms energy optimization in of prediction flips (e.g., better model robustness). However, this difference is again not statistically significant ($p$-value$>$0.5) for both tasks. 

Finally, when looking at sustainability and energy use, improvement results are mixed, depending on the task. 
For Clone Detection (where FLOPs did not show a statistical correlation with energy), the results show that the energy-optimized approach produces models with higher GigaFLOPs ($21\%$) than the FLOPs-optimized approach. 
This difference is statistically significant ($\text{\textit{p}-value}<0.01$) with a medium effect size ($\hat{A}_{12}=0.68$).
Nevertheless, we observe that this increase in FLOPs does not translate to higher energy consumption, and the energy optimization approach outperforms the original approach with a difference of $39 \%$ in median inference energy.
This is a statistically significant difference ($\text{\textit{p}-value}<0.01$) with a large effect size ($\hat{A}_{12}=0.925$).

For Vulnerability Prediction—where FLOPs correlated with energy—the energy-based approach yields models with $36\%$ more GFLOPs and about $30\%$ more energy, though the differences are not statistically significant ($p$-value$>$0.2).

These results suggest that the energy optimization approach can be useful when FLOPs are unreliable or lack a statistical correlation with energy consumption, as is the case for Clone Detection.
For Vulnerability Prediction, the energy approach neither improves nor shows a statistically significant degradation.

\answer{2}{Using surrogate energy instead of FLOPs reduces median inference energy in Clone Detection by $39\%$ (from $5355.83$ J to $3604.50$ J), and this gain is statistically significant ($p<0.01$, $\hat{A}_{12}=0.925$). In Vulnerability Prediction, where FLOPs already correlate with energy, the energy-based optimization changes median energy by $+30\%$ and GFLOPs by $+36\%$, but these differences are not statistically significant ($p>0.2$), and accuracy remains non-significantly different as well ($p>0.3$).}

\subsection{RQ3: Application to generation tasks}
 
\begin{figure*}[ht]
    \centering
    \begin{subfigure}[b]{0.45\textwidth}
        \centering
        \includegraphics[width=\textwidth]{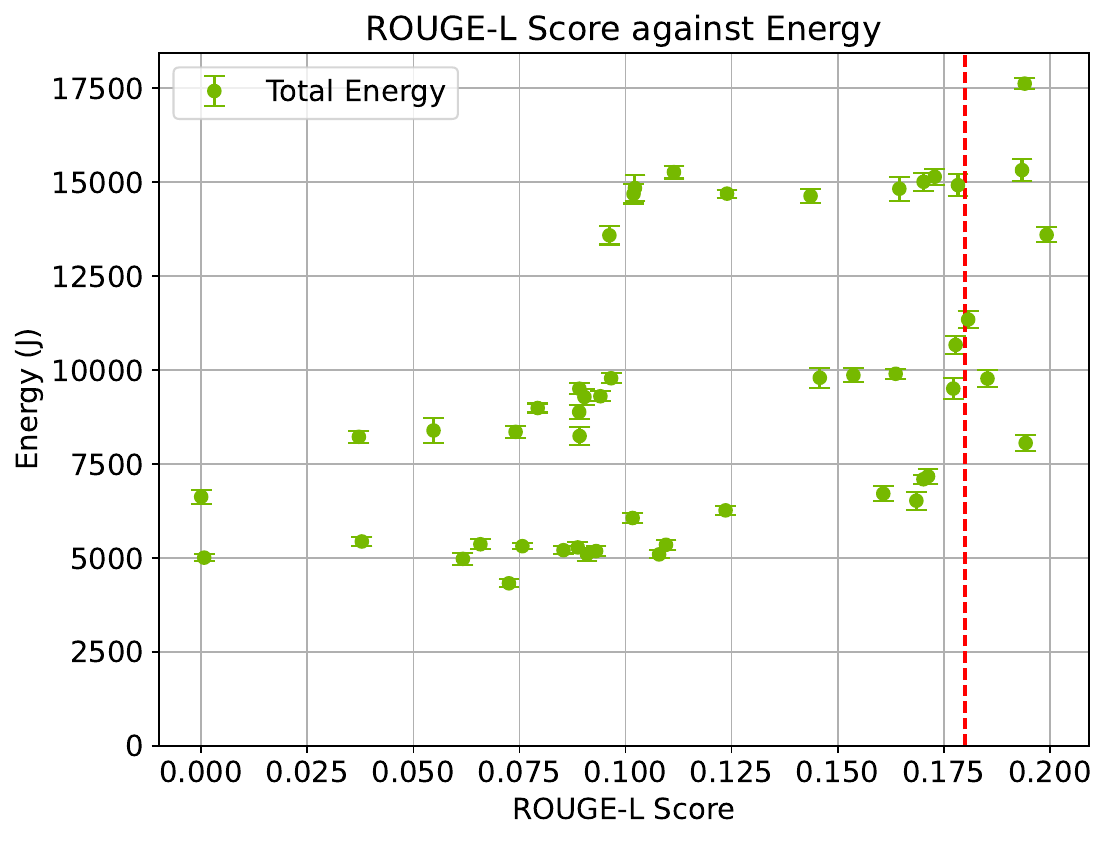}
        \caption{Energy}
        \label{fig:CT5_ROUGE_Energy}
    \end{subfigure}
    \hfill
    \begin{subfigure}[b]{0.45\textwidth}
        \centering
        \includegraphics[width=\textwidth]{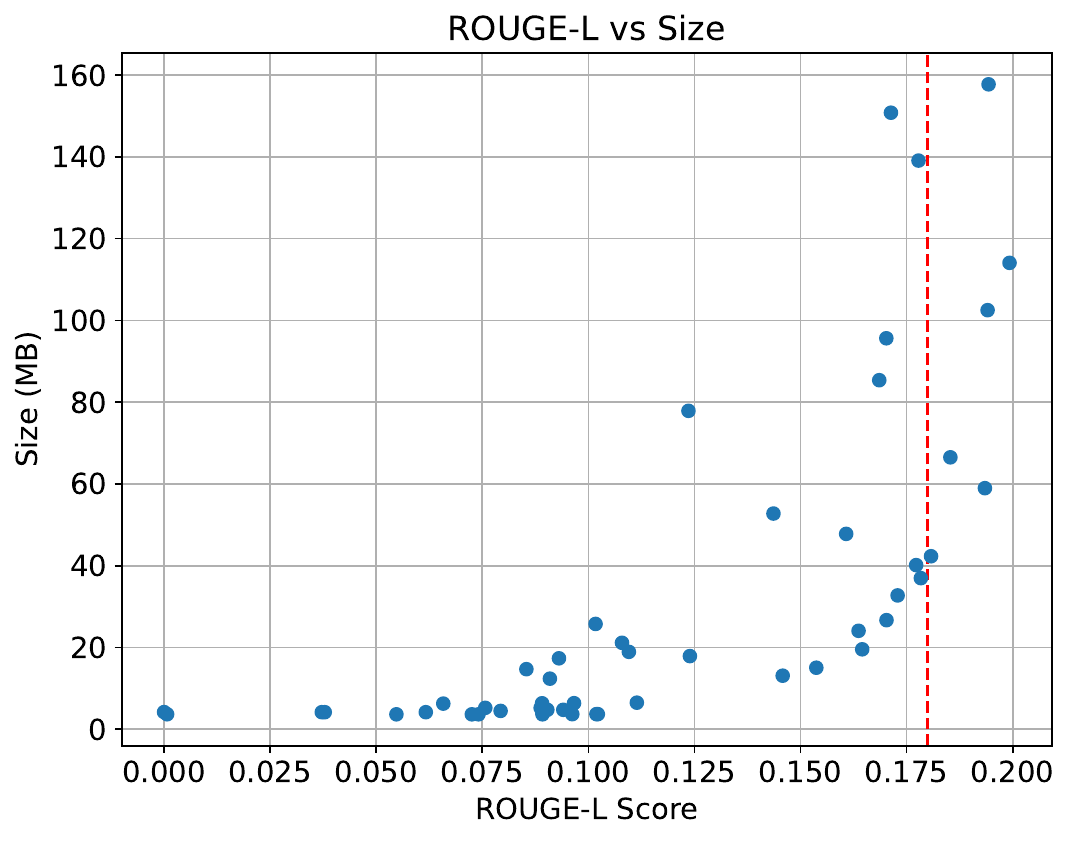}
        \caption{Size}
        \label{fig:CT5_ROUGE_Size}
    \end{subfigure}
    \caption{ROUGE-L of the Pareto front compared to inference energy and model size}
    \label{fig:CT5_Energy_Size}
\end{figure*}

Figure \ref{fig:CT5_Energy_Size} shows the ROUGE-L scores of different generative code summarization models on the Pareto front, along with the other two optimization objectives: energy (Figure \ref{fig:CT5_ROUGE_Energy}) and size (Figure \ref{fig:CT5_ROUGE_Size}).
The red line indicates a threshold for the ROUGE-L score of 0.18. 
By manually inspecting individual amples and generating summaries, and observing that summaries with approximately 0.15 still fit well to the source code, despite the summary being worded differently and being much shorter.
To be conservative with our choice, we selected 0.18 as an acceptable lower bound for our use case.
We observe that, for Code Summarization, the Pareto front shows more constraints and much higher trade-offs between the objectives.
One of the main observations we can make is that, unlike classification tasks, the 3MB selection rule from the previous study \cite{11029766} cannot be applied, since models below 10MB achieve ROUGE-L scores below 0.1, which has a significant impact on performance.

\begin{table}[ht]
\centering
\caption{ROUGE-L, Size and Inference Energy for the teacher model and the most accurate student model. The best result for each metric is highlighted in grey.}
\label{tab:RQ2_Results}
\small
\begin{tabular}{lccc}
\hline
\textbf{Model} & \textbf{ROUGE-L} & \textbf{Size (MB)} & \textbf{Energy (J)} \\
\hline
Teacher & \cellcolor{lightgray} 0.229 & 830 & 82 396 \\
Most Accurate & 0.199 (-13\%) & \cellcolor{lightgray}114 (-86\%) & 13 600 (-83\%) \\
Most Efficient & 0.194 (-15\%) & 157 (-81\%) & \cellcolor{lightgray} 8 059 (-90\%) \\
\hline
\end{tabular}
\end{table}

Table \ref{tab:RQ2_Results} shows the objectives for the teacher model, which obtains an ROUGE-L score of 0.231.
Compared with the most accurate student model, the score is 0.199 (a 13\% decrease).
Therefore, even in the best case, we observe a slight decrease in accuracy.
However, by manually inspecting the output of different models, we see that summaries are mostly coherent.
Student models struggle more with code that uses specialized nomenclature, such as specific libraries or services.
Despite a small decrease in accuracy, this student model achieves an 83\% reduction in energy usage while processing the same number of inference samples and a 86\% reduction in memory footprint.

Other student models can also achieve a similar level of accuracy to the most accurate student while achieving higher energy savings. To illustrate this, Table \ref{tab:RQ2_Results} also showcases the results for the most efficient model obtained under the constraint that the ROUGE-L score exceeds $0.18$.
We defined this threshold based on the performance we observed during manual inspection, since models with this score still perform well with simpler pieces of code.
Then, we selected the most efficient model by dividing the ROUGE-L score by the total inference energy, obtaining an Accuracy per unit of Energy metric.
We selected the model with the highest value in this metric to prioritize both accuracy and efficiency: choose a model that provides greater accuracy with less energy usage. 
This kind of selection criterion can be useful for systems with high energy constraints \cite{efficiencyRatio, 11179880}.
For this model, we observe an even higher energy reduction, up to 90\% compared to the teacher model and 40\% less than the most accurate student, albeit with a slightly higher memory footprint.

\answer{3}{For code summarization, many-objective distillation reduces inference energy from $82{,}396$ J to $8{,}059$ J (most efficient student), a $90\%$ reduction, and reduces model size from $830$ MB to $114$ MB for the most accurate student, an $86\%$ reduction. The performance trade-off is moderate: ROUGE-L drops from $0.229$ to $0.199$ ($-13\%$) for the most accurate student and to $0.194$ ($-15\%$) for the most efficient student.}

\section{Discussion}


\subsection{FLOPs as a proxy for energy}
Our experiments for $\textbf{RQ}_{1}$ show that FLOPs are not a consistent proxy for energy consumption: there is no correlation between FLOPs and inference energy for the Clone Detection task, but a positive correlation for the Vulnerability Prediction task. 
Furthermore, even for the Vulnerability Prediction task, where statistical correlation is present, we still cannot detect a direct relationship between FLOPs and energy consumption.
In this task, we observe some models with similar FLOPs that use vastly different amounts of energy, and models with similar energy that differ widely in FLOPs.
By using FLOPs alone to select an energy-efficient model, an AI engineer would miss models that stand out from the correlation and show a reduced energy consumption despite the elevated number of FLOPs.

These results suggest that FLOPs, as a metric, do not capture all the complexities and components involved in a model's total energy consumption. 
The approach used to compute FLOPs for a model~\cite{DBLP:journals/corr/abs-2003-10555} treats FLOPs as mathematical operations rather than machine instructions, treating operations such as multiplications and additions as equally costly.
A similar thing happens with profilers from libraries, including PyTorch, which use formulas to estimate mathematical operations and, in some cases, ignore operations such as additions.


This computation also uses only a subset of the model's hyperparameters (Hidden Size, Number of Layers, Sequence Length, Vocabulary Size, Intermediate Size, and Number of Heads).
Other factors, such as additional model hyperparameters, dataset characteristics, or inference steps, may influence energy consumption across the whole task.
For example, from the ANOVA test, we observe that the Hidden Activation function also has a statistical correlation with energy.
Each possible activation function has a different implementation, which can affect FLOPs and energy in different ways.
These aspects make FLOPs an unreliable proxy, since the computation understates some aspects and fails to account for others that can affect energy consumption.

The FLOPs computation is also blind to software or hardware optimizations introduced by the libraries or the GPU during execution time.
Underlying libraries provided by the operating system and hardware drivers, along with their versions, can affect the energy consumption of any program \cite{DBLP:conf/icse/RoqueCD25}.
An example of this is the omission of operations due to sparsity, where GPUs introduce optimizations to avoid multiplying by 0.
Something similar can happen with respect to sequence padding.
In this methodology, FLOPs are computed using a maximum sequence length. 
However, not all inputs to the model use all this sequence length, with shorter inputs requiring fewer FLOPs.
Additionally, when processing a batch through an LLM model with a defined batch size, all inputs get padded to the length of the longest input.
The largest sequence sets the length for the entire batch. 
Other sequences are padded with padding tokens, which are ignored during computation using an attention mask.
The FLOPs costs for these padding tokens cannot be assumed to be the same as for normal tokens, since the GPU can introduce optimizations to avoid computations with these tokens.
Therefore, the FLOPs count provided by this methodology is an upper bound rather than an exact count of computations per input.

These optimizations might explain the lack of correlation in the Clone Detection task.
To perform this task, the maximum sequence length is used when tokenizing each code block.
To compare two blocks of code, they are concatenated, giving an effective maximum length of twice the input to the tokenizer.
While this is considered for FLOP computation, it also means the average padding length doubles, resulting in the Clone Detection task having more ``empty'' FLOPs.

In conclusion, FLOPs metrics, as used in this approach and in most of the literature, serve as an upper bound that does not account for all characteristics of the model and task. For simpler tasks or datasets, such as Clone Detection, the actual computation can fall far short of this upper bound, making FLOPs a poor proxy for energy consumption.
In comparison, more complex tasks, such as Vulnerability Prediction, are closer to the upper bound on FLOPs and exhibit higher correlation.

A more accurate proxy for energy consumption than FLOPs is inference time, which shows a strong correlation with energy consumption, according to previous studies \cite{flopsCriticism}.
Compared to energy, this is a much more accessible metric, since it does not require the use of profilers or energy measurement guidelines ~\cite{energyMeasurementsGuide}.
However, to compare with FLOPs, the model still needs to be run.
Running and measuring energy for a model can be extremely time-consuming, especially in many-objective optimization problems, where multiple configurations must be evaluated.
Using inference time is a reasonable tradeoff, since it still requires running the model, but does not require additional setup for energy measurements.
Additionally, inference time provides no insights into the model's energy impact.





From the results for $\textbf{RQ}_{2}$, we learn that, in cases where FLOPs and energy present a correlation (Vulnerability Prediction), using surrogate models instead of FLOPs does not have a statistically significant effect on the final result, for better or worse.
However, when FLOPs cannot accurately reflect energy consumption, the surrogate models bridge this gap and produce more efficient models without impact on accuracy.
These results show that energy surrogate models are preferable to FLOPs, as they can account for more factors that affect sustainability and perform well regardless of whether a correlation is present.

To build the energy models, we need to measure the inference energy consumption of all the sampled models in the surrogate step of \morph{}. 
This adds an additional step to the sampling, training, and evaluating pipeline from the original methodology.
Energy measurements are a costly step, especially in larger student models that tackle more complex tasks, as in the case of CodeT5+ and Code Summarization.
Therefore, it is preferable to use FLOPs when we are certain of the correlation, such as in the Vulnerability Prediction task, since these additional costs do not yield any benefits.

In any case, the trade-offs between FLOPs and energy only apply during the optimization phase of this methodology.
Once \morph{} provides the Pareto front of the optimal models, we observe that, statistical correlations aside, there is no direct relationship between FLOPs and energy.
This is an important consideration when selecting a final student model that meets our constraints.
If we wish to select a model with the lowest possible energy consumption, selecting the one with the lowest FLOPs is not guaranteed to yield the lowest energy consumption.
FLOPs might be useful for solving the optimization problem, as they are easy to compute (and may correlate with energy).
Nevertheless, we propose that developers prioritize actual energy measurements over FLOPs when selecting a final student model, particularly for heavily constrained systems. This recommendation stems from our findings, which indicate that FLOPs are not a reliable proxy for energy consumption in this context, and that related literature should be cautious in its use.


\subsection{Many-Objective Optimization for Generation Tasks}
Our experiments for $\textbf{RQ}_{3}$ show that Many-Objective distillation can be used for generative SE tasks like Code Summarization, albeit with some caveats. 
This technique produces student models with a significant reduction in size (up to 86\% of the teacher model) and energy consumption during inference (up to 90\%).
However, none of the student models match the teacher model's accuracy, resulting in a slight decrease in ROUGE-L score.
This drop is similar to results from the literature on other distillation approaches for summarization~\cite{DBLP:journals/corr/abs-2010-13002, DBLP:journals/ase/SuM24}.

Our study also shows how sensitive state-of-the-art models are when applied to datasets other than their original evaluation dataset.
To mitigate contamination, we are using The Heap \cite{DBLP:conf/forge/KatzyPDI25} as our evaluation dataset.
This dataset follows more strict deduplication techniques than other publicly available datasets to avoid contamination.
This affects final ROUGE-L scores for both teachers and students, with both obtaining scores slightly lower than those of other state-of-the-art summarization models.
For example, our fine-tuned teacher model has an average ROUGE-L of 0.22 in The Heap.
Comparatively, when evaluating the teacher model on CodeXGLUE as in the original CodeT5+ paper \cite{DBLP:conf/emnlp/WangLGB0H23}, we obtain an average ROUGE-L of 0.33, closer to the original results.
This indicates that the model is not generalizing well for a different evaluation dataset.
This calls for more research on how these models should be evaluated and highlights the need for metrics that better reflect the actual correctness of generative model outputs for software engineering problems, as well as for improved generalization beyond predefined benchmarks.


In our experiments, the student model's savings come at the cost of a drop in performance. 
While the observed drop is moderate, it will depend on the use case whether it is acceptable.
For example, local models running on a laptop and used for commit message generation can have a strict energy cap, and the accuracy drop is acceptable for such low-critical tasks.
On the other hand, a model intended to summarize pull request contributions requires an accurate summary to guide the reviewer who will eventually accept or reject the PR.
What is an acceptable drop in accuracy and energy savings will depend on the specific use case, and is hard to define in a research setting.

The Many-objective approach also provides greater flexibility in the student architecture, thanks to weight subcloning.
Other approaches in the literature need a student model that has been pretrained and has some general knowledge first \cite{DBLP:journals/corr/abs-2010-13002, DBLP:journals/ase/SuM24}, or distill into pretrained checkpoints of smaller models.
In this case, the available configurations are only those that are pretrained and open source.
In comparison, our approach copies the most important weights from the teacher model using weight subcloning.
This distillation technique can provide a wider range of hyperparameters while avoiding the expense of pretraining.




\section{Threats to Validity}

Our approach has some limitations regarding its validity. 
First is the distillation loss used for the Code Summarization task. In this paper, we used a standard distillation loss function based on the KL-divergence and the cross-entropy function.
Model distillation is a rapidly developing subfield of AI. Newer methods are being proposed and bring improved and more complex losses \cite{DBLP:journals/corr/abs-2411-06839}, or different distillation paradigms \cite{10.1145/3701551.3703577}.
However, the loss we applied for this paper is widely adopted in state-of-the-art solutions. 
Moreover, the focus of this paper is the application of Many-Objective Problem optimization towards exploration of the configuration space.
We decided to use a standard, well-founded distillation loss and reduce the influence of other variables.
Combining newer methods with Many-Objective solvers might yield improved distillation results.

Additionally, for the generation task, we used the ROUGE-L score as the model's objective, which measures n-gram overlap. 
This metric is widely adopted in state-of-the-art evaluation, but it comes with its own limitations. ROUGE-L overlooks semantic similarity: there can be different valid summaries paraphrased in different ways.
Evaluating semantic similarity would require human judgment or that of a reliable LLM, which incurs significant time and financial costs due to the need for an additional LLM or human participants and does not guarantee the validity of the judgment, especially in the case of LLM-as-a-judge. 
Given the amount of sampling necessary for the method applied in this paper, this kind of metric would be too costly and not entirely objective for the purpose of Many-Objective optimization.
We selected ROUGE-L because, despite its shortcomings, it is widely used in the literature and aligns more closely with human judgment, according to the latest studies~\cite{EVTIKHIEV2023111741}.

\section{Conclusion}
This study investigates the validity of FLOPs as an energy proxy for many-objective LLM knowledge distillation. 
Our findings reveal that the correlation between FLOPs and inference energy of the student model is often unreliable.
For Clone Detection, FLOPs do not present a statistical correlation with energy, while Vulnerability Prediction does.
Besides this correlation, FLOPs does not show a direct relationship with energy.
Selecting the model with lower FLOPs does not always imply selecting the model with lower energy usage.

To address this, we proposed using surrogate energy models to estimate energy consumption for a given hyperparameter set before training.
This approach converges to models that are 39\% more energy-efficient for Clone Detection tasks, bridging the gap in the correlation between FLOPs and energy efficiency.

We also applied this approach to Code Summarization in CodeT5+, testing its potential for generation tasks.
The results showed significant compression levels for students (up to 86\% smaller) and reduced energy usage (up to 90\%), with a modest impact on accuracy.

Future work will expand the generalizability of these findings by applying them to tasks such as code generation.
We also intend to analyze the pitfalls of FLOPs and define metrics that correlate better with energy while requiring less effort than direct energy measurement.

\section{Data Availability}
A replication package is available in Zenodo\footnote{\url{https://doi.org/10.5281/zenodo.20270810}}.
This package contains the scripts to distill surrogate and final models, as well as a script to measure the energy consumption of the models. 
It also contains the data collected from our hardware setup, as well as the analysis scripts used to generate the plots and tables in this paper.

\bibliography{references}

@INPROCEEDINGS{11029766,
  author={Panichella, Annibale},
  booktitle={2025 IEEE/ACM 47th International Conference on Software Engineering (ICSE)}, 
  title={Metamorphic-Based Many-Objective Distillation of LLMs for Code-Related Tasks}, 
  year={2025},
  volume={},
  number={},
  xpages={1001-1013},
  doi={10.1109/ICSE55347.2025.00230},
  ISSN={1558-1225},
  month={April},}

@inproceedings{10.1145/3321707.3321839,
author = {Panichella, Annibale},
title = {An adaptive evolutionary algorithm based on non-euclidean geometry for many-objective optimization},
year = {2019},
isbn = {9781450361118},
xpublisher = {Association for Computing Machinery},
xaddress = {New York, NY, USA},
xurl = {https://doi.org/10.1145/3321707.3321839},
doi = {10.1145/3321707.3321839},
booktitle = {Genetic and Evolutionary Computation Conference},
xpages = {595–603},
xnumpages = {9},
xlocation = {Prague, Czech Republic},
series = {GECCO '19}
}

@inproceedings{DBLP:conf/emnlp/WangLGB0H23,
  author       = {Yue Wang and
                  Hung Le and
                  Akhilesh Gotmare and
                  Nghi D. Q. Bui and
                  others},
  title        = {CodeT5+: Open Code Large Language Models for Code Understanding and
                  Generation},
  booktitle    = {Conference on Empirical Methods in Natural
                  Language Processing},
  xpages        = {1069--1088},
  xpublisher    = {Association for Computational Linguistics},
  year         = {2023},
  xurl          = {https://doi.org/10.18653/v1/2023.emnlp-main.68},
  doi          = {10.18653/V1/2023.EMNLP-MAIN.68},
  bibsource    = {dblp computer science bibliography, https://dblp.org}
}

@inproceedings{10.1145/3324884.3416591,
author = {Liu, Fang and Li, Ge and Zhao, Yunfei and Jin, Zhi},
title = {Multi-task learning based pre-trained language model for code completion},
year = {2021},
isbn = {9781450367684},
publisher = {Association for Computing Machinery},
address = {New York, NY, USA},
url = {https://doi.org/10.1145/3324884.3416591},
doi = {10.1145/3324884.3416591},
booktitle = {Proceedings of the 35th IEEE/ACM International Conference on Automated Software Engineering},
pages = {473–485},
numpages = {13},
location = {Virtual Event, Australia},
series = {ASE '20}
}

@inproceedings{10.5555/3618408.3618894,
author = {Guo, Daya and Xu, Canwen and Duan, Nan and Yin, Jian and others},
title = {LongCoder: a long-range pre-trained language model for code completion},
year = {2023},
xpublisher = {JMLR.org},
booktitle = {40th International Conference on Machine Learning},
xarticleno = {486},
xnumpages = {10},
location = {Honolulu, Hawaii, USA},
series = {ICML'23}
}

@INPROCEEDINGS{10556419,
  author={Khajezade, Mohamad and Wu, Jie Jw and Fard, Fatemeh Hendijani and Rodríguez-Pérez, Gema and others},
  booktitle={32nd International Conference on Program Comprehension (ICPC)}, 
  title={Investigating the Efficacy of Large Language Models for Code Clone Detection}, 
  year={2024},
  volume={},
  number={},
  xpages={161-165},
  doi={10.1145/3643916.364503}
  }

@inproceedings{10.1145/3639476.3639762,
author = {Zhou, Xin and Zhang, Ting and Lo, David},
title = {Large Language Model for Vulnerability Detection: Emerging Results and Future Directions},
year = {2024},
isbn = {9798400705007},
xpublisher = {Association for Computing Machinery},
xaddress = {New York, NY, USA},
xurl = {https://doi.org/10.1145/3639476.3639762},
doi = {10.1145/3639476.3639762},
booktitle = {44th International Conference on Software Engineering: New Ideas and Emerging Results},
xpages = {47–51},
xnumpages = {5},
location = {Lisbon, Portugal},
series = {ICSE-NIER'24}
}

@article{DBLP:journals/corr/abs-2102-04664,
  author       = {Shuai Lu and
                  Daya Guo and
                  Shuo Ren and
                  Junjie Huang and
                  others},
  title        = {CodeXGLUE: {A} Machine Learning Benchmark Dataset for Code Understanding
                  and Generation},
  journal      = {CoRR},
  xvolume       = {abs/2102.04664},
  year         = {2021},
 xurl          = {https://arxiv.org/abs/2102.04664},
  eprinttype    = {arXiv},
  eprint       = {2102.04664},
  xbiburl       = {https://dblp.org/rec/journals/corr/abs-2102-04664.bib},
  xbibsource    = {dblp computer science bibliography, https://dblp.org}
}

@misc{energyMeasurementsGuide,
    author = {Luís Cruz},
    title = {Green Software Engineering Done Right: a Scientific Guide to Set Up Energy Efficiency Experiments},
    url = {http://luiscruz.github.io/2021/10/10/scientific-guide.html},
    xurldate = {2024-05-27},
    year = {2021}
}

@misc{sallou2023energibridge,
      title={EnergiBridge: Empowering Software Sustainability through Cross-Platform Energy Measurement}, 
      author={June Sallou and Luís Cruz and Thomas Durieux},
      year={2023},
      eprint={2312.13897},
      archivePrefix={arXiv},
      primaryClass={cs.SE}
}

@inproceedings{DBLP:conf/icse/RoqueCD25,
  author       = {Enrique Barba Roque and
                  Luis Cruz and
                  Thomas Durieux},
  title        = {Unveiling the Energy Vampires: {A} Methodology for Debugging Software
                  Energy Consumption},
  booktitle    = {47th {IEEE/ACM} International Conference on Software Engineering,
                  {ICSE}, Ottawa},
  xpages        = {2406--2418},
  xpublisher    = {{IEEE}},
  year         = {2025},
 xurl          = {https://doi.org/10.1109/ICSE55347.2025.00118},
  doi          = {10.1109/ICSE55347.2025.00118},
  bibsource    = {dblp computer science bibliography, https://dblp.org}
}

@inproceedings{DBLP:conf/mlsys/WuRGAAMCBHBGGOM22,
  author       = {Carole{-}Jean Wu and
                  Ramya Raghavendra and
                  Udit Gupta and
                  Bilge Acun and
                  others},
  title        = {Sustainable {AI:} Environmental Implications, Challenges and Opportunities},
  booktitle    = {5th Conference on Machine Learning and Systems,
                  MLSys 2022, Santa Clara},
  xpublisher    = {mlsys.org},
  year         = {2022},
  xurl          = {https://proceedings.mlsys.org/paper\_files/paper/2022/hash/462211f67c7d858f663355eff93b745e-Abstract.html},
  bibsource    = {dblp computer science bibliography, https://dblp.org}
}

@article{10.1145/3381831,
author = {Schwartz, Roy and Dodge, Jesse and Smith, Noah A. and Etzioni, Oren},
title = {Green AI},
year = {2020},
issue_date = {December 2020},
publisher = {Association for Computing Machinery},
address = {New York, NY, USA},
volume = {63},
number = {12},
issn = {0001-0782},
url = {https://doi.org/10.1145/3381831},
doi = {10.1145/3381831},
journal = {Commun. ACM},
month = nov,
pages = {54–63},
numpages = {10}
}

@inproceedings{10.1145/3551349.3556964,
author = {Shi, Jieke and Yang, Zhou and Xu, Bowen and Kang, Hong Jin and others},
title = {Compressing Pre-trained Models of Code into 3 MB},
year = {2023},
isbn = {9781450394758},
xpublisher = {Association for Computing Machinery},
xaddress = {New York, NY, USA},
xurl = {https://doi.org/10.1145/3551349.3556964},
doi = {10.1145/3551349.3556964},
booktitle = {37th IEEE/ACM International Conference on Automated Software Engineering},
xarticleno = {24},
xnumpages = {12},
location = {Rochester, MI, USA},
series = {ASE '22}
}

@inproceedings{10.1145/3639475.3640097,
author = {Shi, Jieke and Yang, Zhou and Kang, Hong Jin and Xu, Bowen and others},
title = {Greening Large Language Models of Code},
year = {2024},
isbn = {9798400704994},
xpublisher = {Association for Computing Machinery},
xaddress = {New York, NY, USA},
xurl = {https://doi.org/10.1145/3639475.3640097},
doi = {10.1145/3639475.3640097},
booktitle = {46th International Conference on Software Engineering: Software Engineering in Society},
xpages = {142–153},
xnumpages = {12},
location = {Lisbon, Portugal},
series = {ICSE-SEIS'24}
}

@INPROCEEDINGS{9463109,
  author={Svyatkovskiy, Alexey and Lee, Sebastian and Hadjitofi, Anna and Riechert, Maik and others},
  booktitle={2021 IEEE/ACM 18th International Conference on Mining Software Repositories (MSR)}, 
  title={Fast and Memory-Efficient Neural Code Completion}, 
  year={2021},
  volume={},
  number={},
  xpages={329-340},
  doi={10.1109/MSR52588.2021.00045}}

@article{DBLP:journals/corr/abs-2312-09299,
  author       = {Mohammad Samragh and
                  Mehrdad Farajtabar and
                  Sachin Mehta and
                  Raviteja Vemulapalli and
                  others},
  title        = {Weight subcloning: direct initialization of transformers using larger
                  pretrained ones},
  journal      = {CoRR},
  xvolume       = {abs/2312.09299},
  year         = {2023},
 xurl          = {https://doi.org/10.48550/arXiv.2312.09299},
  doi          = {10.48550/ARXIV.2312.09299},
  xeprinttype    = {arXiv},
  xeprint       = {2312.09299},
  bibsource    = {dblp computer science bibliography, https://dblp.org}
}

@inproceedings{DBLP:conf/acl/AhmadCRC20,
  author       = {Wasi Uddin Ahmad and
                  Saikat Chakraborty and
                  Baishakhi Ray and
                  Kai{-}Wei Chang},
  title        = {A Transformer-based Approach for Source Code Summarization},
  booktitle    = {58th Annual Meeting of the Association for Computational
                  Linguistics, {ACL}},
  xpages        = {4998--5007},
  xpublisher    = {Association for Computational Linguistics},
  year         = {2020},
 xurl          = {https://doi.org/10.18653/v1/2020.acl-main.449},
  doi          = {10.18653/V1/2020.ACL-MAIN.449},
  bibsource    = {dblp computer science bibliography, https://dblp.org}
}

@inproceedings{10.1145/3630106.3658542,
author = {Luccioni, Sasha and Jernite, Yacine and Strubell, Emma},
title = {Power Hungry Processing: Watts Driving the Cost of AI Deployment?},
year = {2024},
xisbn = {9798400704505},
xpublisher = {Association for Computing Machinery},
xaddress = {New York, NY, USA},
xurl = {https://doi.org/10.1145/3630106.3658542},
doi = {10.1145/3630106.3658542},
booktitle = {Conference on Fairness, Accountability, and Transparency},
xpages = {85–99},
xnumpages = {15},
location = {Rio de Janeiro, Brazil},
series = {FAccT '24}
}

@article{DBLP:journals/corr/abs-2010-13002,
  author       = {Sam Shleifer and
                  Alexander M. Rush},
  title        = {Pre-trained Summarization Distillation},
  journal      = {CoRR},
  volume       = {abs/2010.13002},
  year         = {2020},
  xurl          = {https://arxiv.org/abs/2010.13002},
  eprinttype    = {arXiv},
  eprint       = {2010.13002},
  bibsource    = {dblp computer science bibliography, https://dblp.org}
}

@article{DBLP:journals/ase/SuM24,
  author       = {Chia{-}Yi Su and
                  Collin McMillan},
  title        = {Distilled {GPT} for source code summarization},
  journal      = {Autom. Softw. Eng.},
  volume       = {31},
  number       = {1},
  pages        = {22},
  year         = {2024},
 xurl          = {https://doi.org/10.1007/s10515-024-00421-4},
  doi          = {10.1007/S10515-024-00421-4},
  bibsource    = {dblp computer science bibliography, https://dblp.org}
}

@article{DBLP:journals/corr/abs-2411-06839,
  author       = {Runming Yang and
                  Taiqiang Wu and
                  Jiahao Wang and
                  Pengfei Hu and
                  Ngai Wong and
                  Yujiu Yang},
  title        = {LLM-Neo: Parameter Efficient Knowledge Distillation for Large Language
                  Models},
  journal      = {CoRR},
  volume       = {abs/2411.06839},
  year         = {2024},
 xurl          = {https://doi.org/10.48550/arXiv.2411.06839},
  doi          = {10.48550/ARXIV.2411.06839},
  xeprinttype    = {arXiv},
  xeprint       = {2411.06839},
  xbiburl       = {https://dblp.org/rec/journals/corr/abs-2411-06839.bib},
  xbibsource    = {dblp computer science bibliography, https://dblp.org}
}

@inproceedings{10.1145/3701551.3703577,
author = {Tian, Yijun and Han, Yikun and Chen, Xiusi and Wang, Wei and Chawla, Nitesh V.},
title = {Beyond Answers: Transferring Reasoning Capabilities to Smaller LLMs Using Multi-Teacher Knowledge Distillation},
year = {2025},
isbn = {9798400713293},
xpublisher = {Association for Computing Machinery},
xaddress = {New York, NY, USA},
xurl = {https://doi.org/10.1145/3701551.3703577},
doi = {10.1145/3701551.3703577},
booktitle = {18th ACM International Conference on Web Search and Data Mining},
xpages = {251–260},
xnumpages = {10},
location = {Hannover, Germany},
series = {WSDM '25}
}

@techreport{IEA2025EnergyAI,
  title        = {Energy and AI},
  author       = {{International Energy Agency}},
  institution  = {International Energy Agency (IEA)},
  address      = {Paris, France},
  year         = {2025},
  month        = apr,
  type         = {Special Report},
  url          = {https://www.iea.org/reports/energy-and-ai},
  note         = {Published April 2025. CC BY 4.0 licence},
}

@misc{hinton2015distillingknowledgeneuralnetwork,
      title={Distilling the Knowledge in a Neural Network}, 
      author={Geoffrey Hinton and Oriol Vinyals and Jeff Dean},
      year={2015},
      eprint={1503.02531},
      archivePrefix={arXiv},
      primaryClass={stat.ML},
      xurl={https://arxiv.org/abs/1503.02531}, 
}

@article{flopsToEnergy,
title = {Trends in AI inference energy consumption: Beyond the performance-vs-parameter laws of deep learning},
journal = {Sustainable Computing: Informatics and Systems},
xvolume = {38},
xpages = {100857},
year = {2023},
issn = {2210-5379},
doi = {https://doi.org/10.1016/j.suscom.2023.100857},
xurl = {https://www.sciencedirect.com/science/article/pii/S2210537923000124},
author = {Radosvet Desislavov and Fernando Martínez-Plumed and José Hernández-Orallo}
}

@inproceedings{flopsCriticism,
  author       = {Andrea Asperti and
                  Davide Evangelista and
                  Moreno Marzolla},
  xeditor       = {Giuseppe Nicosia and
                  Varun Ojha and
                  Emanuele La Malfa and
                  Gabriele La Malfa and
                  Giorgio Jansen and
                  Panos M. Pardalos and
                  Giovanni Giuffrida and
                  Renato Umeton},
  title        = {Dissecting FLOPs Along Input Dimensions for GreenAI Cost Estimations},
  booktitle    = {Machine Learning, Optimization, and Data Science - 7th International
                  Conference, {LOD}, Grasmere},
  xseries       = {Lecture Notes in Computer Science},
  xpages        = {86--100},
  xpublisher    = {Springer},
  year         = {2021},
  xurl          = {https://doi.org/10.1007/978-3-030-95470-3\_7},
  doi          = {10.1007/978-3-030-95470-3\_7},
  xbiburl       = {https://dblp.org/rec/conf/mod/AspertiEM21.bib},
  xbibsource    = {dblp computer science bibliography, https://dblp.org}
}

@misc{lozhkov2024starcoder,
      title={StarCoder 2 and The Stack v2: The Next Generation}, 
      author={Anton Lozhkov and Raymond Li and Loubna Ben Allal and others},
      year={2024},
      eprint={2402.19173},
      archivePrefix={arXiv},
      primaryClass={cs.SE}
}

@INPROCEEDINGS{6976121,
  author={Svajlenko, Jeffrey and Islam, Judith F. and Keivanloo, Iman and Roy, Chanchal K. and others},
  booktitle={2014 IEEE International Conference on Software Maintenance and Evolution}, 
  title={Towards a Big Data Curated Benchmark of Inter-project Code Clones}, 
  year={2014},
  volume={},
  number={},
  xpages={476-480},
  doi={10.1109/ICSME.2014.77}}

@inproceedings{DBLP:conf/nips/ZhouLSD019,
author = {Zhou, Yaqin and Liu, Shangqing and Siow, Jingkai and Du, Xiaoning and others},
title = {Devign: effective vulnerability identification by learning comprehensive program semantics via graph neural networks},
url = {https://dl.acm.org/doi/10.5555/3454287.3455202},
year = {2019},
xpublisher = {Curran Associates Inc.},
xaddress = {Red Hook, NY, USA},
booktitle = {33rd International Conference on Neural Information Processing Systems},
xarticleno = {915},
xnumpages = {11}
}

@inproceedings{DBLP:conf/forge/KatzyPDI25,
  author       = {Jonathan Katzy and
                  Razvan Mihai Popescu and
                  Arie van Deursen and
                  Maliheh Izadi},
  title        = {The Heap: {A} Contamination-Free Multilingual Code Dataset for Evaluating
                  Large Language Models},
  booktitle    = {{IEEE/ACM} Second International Conference on {AI} Foundation Models
                  and Software Engineering, Forge@ICSE 2025, Ottawa},
  xpages        = {151--155},
  xpublisher    = {{IEEE}},
  year         = {2025},
 xurl          = {https://doi.org/10.1109/Forge66646.2025.00025},
  doi          = {10.1109/FORGE66646.2025.00025},
  bibsource    = {dblp computer science bibliography, https://dblp.org}
}

@article{DBLP:journals/corr/abs-2009-08366,
  author       = {Daya Guo and
                  Shuo Ren and
                  Shuai Lu and
                  Zhangyin Feng and
                  others},
  title        = {GraphCodeBERT: Pre-training Code Representations with Data Flow},
  journal      = {CoRR},
  volume       = {abs/2009.08366},
  year         = {2020},
 xurl          = {https://arxiv.org/abs/2009.08366},
  eprinttype    = {arXiv},
  eprint       = {2009.08366},
  bibsource    = {dblp computer science bibliography, https://dblp.org}
}

@article{DBLP:journals/corr/abs-2002-08155,
  author       = {Zhangyin Feng and
                  Daya Guo and
                  Duyu Tang and
                  Nan Duan and
                  others},
  title        = {CodeBERT: {A} Pre-Trained Model for Programming and Natural Languages},
  journal      = {CoRR},
  volume       = {abs/2002.08155},
  year         = {2020},
 xurl          = {https://arxiv.org/abs/2002.08155},
  eprinttype    = {arXiv},
  eprint       = {2002.08155},
  bibsource    = {dblp computer science bibliography, https://dblp.org}
}

@article{DBLP:journals/corr/abs-2003-10555,
  author       = {Kevin Clark and
                  Minh{-}Thang Luong and
                  Quoc V. Le and
                  Christopher D. Manning},
  title        = {{ELECTRA:} Pre-training Text Encoders as Discriminators Rather Than Generators},
  journal      = {CoRR},
  xvolume       = {abs/2003.10555},
  year         = {2020},
 xurl          = {https://arxiv.org/abs/2003.10555},
  eprinttype    = {arXiv},
  eprint       = {2003.10555},
  xbiburl       = {https://dblp.org/rec/journals/corr/abs-2003-10555.bib},
  xbibsource    = {dblp computer science bibliography, https://dblp.org}
}

@INPROCEEDINGS {kl1,
author = { Zhao, Borui and Cui, Quan and Song, Renjie and Qiu, Yiyu and Liang, Jiajun },
booktitle = { 2022 IEEE/CVF Conference on Computer Vision and Pattern Recognition (CVPR) },
title = {{ Decoupled Knowledge Distillation }},
year = {2022},
volume = {},
ISSN = {},
xpages = {11943-11952},
doi = {10.1109/CVPR52688.2022.01165},
xurl = {https://doi.ieeecomputersociety.org/10.1109/CVPR52688.2022.01165},
xpublisher = {IEEE Computer Society},
xaddress = {Los Alamitos, CA, USA},
month =Jun}

@inproceedings{kl2,
  author       = {Pengguang Chen and
                  Shu Liu and
                  Hengshuang Zhao and
                  Jiaya Jia},
  title        = {Distilling Knowledge via Knowledge Review},
  booktitle    = {{IEEE} Conference on Computer Vision and Pattern Recognition, {CVPR}
                  2021},
  xpages        = {5008--5017},
  xpublisher    = {Computer Vision Foundation / {IEEE}},
  year         = {2021},
 xurl          = {https://openaccess.thecvf.com/content/CVPR2021/html/Chen\_Distilling\_Knowledge\_via\_Knowledge\_Review\_CVPR\_2021\_paper.html},
  doi          = {10.1109/CVPR46437.2021.00497},
  bibsource    = {dblp computer science bibliography, https://dblp.org}
}

@inproceedings{klsupport,
  author       = {Taiqiang Wu and
                  Chaofan Tao and
                  Jiahao Wang and
                  Runming Yang and others},
  title        = {Rethinking Kullback-Leibler Divergence in Knowledge Distillation for
                  Large Language Models},
  booktitle    = {31st International Conference on Computational
                  Linguistics, {COLING}, Abu Dhabi},
  xpages        = {5737--5755},
  xpublisher    = {Association for Computational Linguistics},
  year         = {2025},
 xurl          = {https://aclanthology.org/2025.coling-main.383/},
  bibsource    = {dblp computer science bibliography, https://dblp.org}
}

@book{wilcoxon,
  address = {New York, NY [u.a.]},
  author = {Conover, {William Jay}},
  edition = {3. ed},
  isbn = {0471160687},
  pagetotal = {VIII, 584},
  ppn_gvk = {24551600X},
  publisher = {Wiley},
  series = {Wiley series in probability and statistics},
  title = {Practical nonparametric statistics},
  xurl = {http://gso.gbv.de/DB=2.1/CMD?ACT=SRCHA&SRT=YOP&IKT=1016&TRM=ppn+24551600X&sourceid=fbw_bibsonomy},
  year = 1999
}

@article{a12,
author = {András Vargha and Harold D. Delaney},
title ={A Critique and Improvement of the CL Common Language Effect Size Statistics of McGraw and Wong},
journal = {Journal of Educational and Behavioral Statistics},
xvolume = {25},
xnumber = {2},
xpages = {101-132},
year = {2000},
doi = {10.3102/10769986025002101},
xURL = {https://doi.org/10.3102/10769986025002101},
xeprint = {https://doi.org/10.3102/10769986025002101}
}

@INPROCEEDINGS{10371661,
  author={del Rey, Santiago and Martínez-Fernández, Silverio and Cruz, Luís and Franch, Xavier},
  booktitle={49th Euromicro Conference on Software Engineering and Advanced Applications (SEAA)}, 
  title={Do DL models and training environments have an impact on energy consumption?}, 
  year={2023},
  volume={},
  number={},
  xpages={150-158},
  doi={10.1109/SEAA60479.2023.00031},
  ISSN={2376-9521},
  month={Sep.},}

@inproceedings{jiao-etal-2020-tinybert,
    title = "{T}iny{BERT}: Distilling {BERT} for Natural Language Understanding",
    author = "Jiao, Xiaoqi  and
      Yin, Yichun  and
      Shang, Lifeng  and
      Jiang, Xin  and
      others",
    booktitle = "Findings of the Association for Computational Linguistics: EMNLP",
    month = nov,
    year = "2020",
    xaddress = "Online",
    xpublisher = "Association for Computational Linguistics",
   xurl = "https://aclanthology.org/2020.findings-emnlp.372/",
    doi = "10.18653/v1/2020.findings-emnlp.372",
    pages = "4163--4174"
}

@inproceedings{liu-etal-2020-fastbert,
    title = "{F}ast{BERT}: a Self-distilling {BERT} with Adaptive Inference Time",
    author = "Liu, Weijie  and
      Zhou, Peng  and
      Wang, Zhiruo  and
      Zhao, Zhe  and
      others",
    booktitle = "58th Annual Meeting of the Association for Computational Linguistics",
    month = jul,
    year = "2020",
    xaddress = "Online",
    xpublisher = "Association for Computational Linguistics",
   xurl = "https://aclanthology.org/2020.acl-main.537/",
    doi = "10.18653/v1/2020.acl-main.537",
    xpages = "6035--6044"
}

@inproceedings{li-etal-2021-accelerating,
    title = "Accelerating {BERT} Inference for Sequence Labeling via Early-Exit",
    author = "Li, Xiaonan  and
      Shao, Yunfan  and
      Sun, Tianxiang  and
      Yan, Hang  and
      others",
    booktitle = "59th Annual Meeting of the Association for Computational Linguistics and the 11th International Joint Conference on Natural Language Processing",
    month = aug,
    year = "2021",
    xaddress = "Online",
    xpublisher = "Association for Computational Linguistics",
   xurl = "https://aclanthology.org/2021.acl-long.16/",
    doi = "10.18653/v1/2021.acl-long.16",
    xpages = "189--199"
}

@article{EVTIKHIEV2023111741,
title = {Out of the BLEU: How should we assess quality of the Code Generation models?},
journal = {Journal of Systems and Software},
volume = {203},
pages = {111741},
year = {2023},
issn = {0164-1212},
doi = {https://doi.org/10.1016/j.jss.2023.111741},
xurl = {https://www.sciencedirect.com/science/article/pii/S016412122300136X},
author = {Mikhail Evtikhiev and Egor Bogomolov and Yaroslav Sokolov and Timofey Bryksin}
}

@misc{deepseekai2025deepseekr1incentivizingreasoningcapability,
      title={DeepSeek-R1: Incentivizing Reasoning Capability in LLMs via Reinforcement Learning}, 
      author={DeepSeek-AI},
      year={2025},
      eprint={2501.12948},
      archivePrefix={arXiv},
      primaryClass={cs.CL},
      xurl={https://arxiv.org/abs/2501.12948}, 
}

@article{efficiencyRatio,
	author = {Cueto-Mendoza, Eduardo and Kelleher, John},
	date = {2024/10/28},
	doi = {10.1007/s10462-024-10943-8},
	id = {Cueto-Mendoza2024},
	isbn = {1573-7462},
	journal = {Artificial Intelligence Review},
	number = {12},
	pages = {349},
	title = {A framework for measuring the training efficiency of a neural architecture},
	url = {https://doi.org/10.1007/s10462-024-10943-8},
	volume = {57},
	year = {2024}}

@INPROCEEDINGS{11179880,
  author={Roque, Enrique Barba and Cruz, Luis},
  booktitle={11th International Conference on ICT for Sustainability (ICT4S)}, 
  title={Energy Aware Development of Neuromorphic Implantables: From Metrics to Action}, 
  year={2025},
  volume={},
  number={},
  xpages={198-208},
  doi={10.1109/ICT4S68164.2025.00028}
  }

@article{DBLP:journals/corr/abs-1910-09700,
  author       = {Alexandre Lacoste and
                  Alexandra Luccioni and
                  Victor Schmidt and
                  Thomas Dandres},
  title        = {Quantifying the Carbon Emissions of Machine Learning},
  journal      = {CoRR},
  volume       = {abs/1910.09700},
  year         = {2019},
 xurl          = {http://arxiv.org/abs/1910.09700},
  eprinttype    = {arXiv},
  eprint       = {1910.09700},
  bibsource    = {dblp computer science bibliography, https://dblp.org}
}

@inproceedings{fan2023large,
  title={Large language models for software engineering: Survey and open problems},
  author={Fan, Angela and Gokkaya, Beliz and Harman, Mark and Lyubarskiy, Mitya and others},
  booktitle={2023 International Conference on Software Engineering: Future of Software Engineering (ICSE-FoSE)},
  xpages={31--53},
  year={2023},
  organization={IEEE},
  doi={10.1109/ICSE-FoSE59343.2023.00008}
}

\end{document}